\documentclass[11pt]{article}

\usepackage[left=1in, right=1in, top=1in, bottom=1in, margin=1in]{geometry}

\usepackage{tcolorbox}
\usepackage{ninecolors}
\usepackage{xcolor}

\usepackage{amsmath,amssymb,xspace,graphicx,relsize,bm,breqn,multirow}
\usepackage{multicol}

\usepackage{braket}
\usepackage{qcircuit}
\usepackage{adjustbox}

\usepackage{float}
\usepackage[linesnumbered,ruled,vlined]{algorithm2e}
\SetKwInput{KwInput}{Input}
\SetKwInput{KwOutput}{Output}
\SetKwInOut{Promise}{Promise}
\SetKwInput{Goal}{Goal}
\SetKwProg{Fn}{function}{}{}
\SetKwFor{RepTimes}{repeat}{times}{}
\SetKwFunction{Ver}{Verify}
\SetKwFunction{Prep}{PrepareState}
\SetKwComment{Comment}{/* }{ */}
\usepackage{algorithmic}

\newtcolorbox{myalgorithm}[1][]{
    colback=gray!10, 
    colframe=black, 
    arc=5pt, 
    boxrule=0.5pt, 
    left=0pt, right=0pt, top=0pt, bottom=0pt 
}
\usepackage[margin=1in]{geometry}

\usepackage{amsthm}
\usepackage{dsfont}
\usepackage{array}
\usepackage{makecell}
\newcommand{\Fe}{\ensuremath{\mathcal{F}}}

\newcommand{\C}{\ensuremath{\mathcal{C}}}

\newcommand{\poly}{\ensuremath{\mathsf{poly}}}

\newcommand{\id}{\ensuremath{\mathbb{I}}}

\usepackage{upgreek}
\usepackage{enumerate}

\usepackage[pagebackref]{hyperref}
\usepackage[usestackEOL]{stackengine}

\usepackage{cleveref}
\renewcommand{\cref}{\Cref}
\usepackage{thm-restate,mathrsfs}
\usepackage{enumerate}
\usepackage{array}
\usepackage{parskip}
\def\01{\{0,1\}}

\newcommand{\ketbra}[2]{|#1\rangle\langle#2|}
\hypersetup{
	colorlinks,
	linkcolor={blue!100!black},
	citecolor={red!100!black},
}

\newcommand{\be}{\begin{equation}}
\newcommand{\ee}{\end{equation}}
\newcommand{\ba}{\begin{array}}
\newcommand{\ea}{\end{array}}
\newcommand{\bea}{\begin{eqnarray}}
\newcommand{\eea}{\end{eqnarray}}

\usepackage{mathtools}

\DeclarePairedDelimiter\floor{\lfloor}{\rfloor}

\DeclareMathOperator{\Tr}{Tr}
\newcommand{\ra}{\rangle}
\newcommand{\la}{\langle}
\newcommand{\sket}[1]{| #1 \rangle\!\rangle}

\newcommand{\opt}{\textsf{opt}}

\newcommand{\norm}[1]{\left\lVert#1\right\rVert}

\newcommand{\calF}{{\cal F }}

\newcommand{\calC}{{\cal C }}
\newcommand{\calS}{{\cal S }}

\newcommand{\calM}{{\cal M }}

\newcommand{\Cliff}{\textsc{Cliff}}
\newcommand{\Stab}{\textsc{Stab}}
\newcommand{\EPR}{\textsf{EPR}}

\newcommand{\dist}{\textsf{dist}}
\newcommand{\op}{\mathrm{op}}

\usepackage{amsthm}

\def\01{\{0,1\}}

\definecolor{citegreen}{HTML}{208054}
\definecolor{citeblue}{HTML}{0055cc}

\NineColors{saturation=high}
\hypersetup{
    breaklinks=true,   
    colorlinks=true, 
    linkcolor=blue3, 
    citecolor=green5, 
    urlcolor=blue3, 
}

\newtheorem{theorem}{Theorem}[section]
\newtheorem{definition}[theorem]{Definition}

\newtheorem{lemma}[theorem]{Lemma}

\newtheorem{corollary}[theorem]{Corollary}

\newtheorem{fact}[theorem]{Fact}
\newtheorem{claim}[theorem]{Claim}

\newtcolorbox{boxeddefinition}[1]{
  colback=blue!5!white,    
  colframe=blue!75!black,  
  fonttitle=\bfseries,
  arc=2mm,                 
  title={Definition \thedefinition: #1},
  before upper={\stepcounter{definition}} 
}

\usepackage{thm-restate,mathrsfs} 

\usepackage{tcolorbox}

\makeatletter
\def\widebreve{\mathpalette\wide@breve}
\def\wide@breve#1#2{\sbox\z@{$#1#2$}%
     \mathop{\vbox{\m@th\ialign{##\crcr
\kern0.08em\brevefill#1{0.8\wd\z@}\crcr\noalign{\nointerlineskip}%
                    $\hss#1#2\hss$\crcr}}}\nolimits}
\def\brevefill#1#2{$\m@th\sbox\tw@{$#1($}%
    \hss\resizebox{#2}{\wd\tw@}{\rotatebox[origin=c]{90}{\upshape(}}\hss$}
\makeatletter

\usepackage{todonotes}
\usepackage{xcolor} 

\title{Learning Clifford-structured quantum unitaries and Hamiltonians}
\author{Arkopal Dutt\thanks{IBM Research} \and  Dale Jacobs$^\dagger$ \and John Jeang$^\dagger$ \and Saeed Mehraban$^\dagger$ \and Vladimir Podolskii\thanks{Tufts University}}

\date{\today{}}

\begin{document}

\maketitle
\begin{abstract}
Learning algorithms for structured quantum unitaries and Hamiltonians have primarily considered classes of processes that are local or sparse in the Pauli basis. We turn our attention to learning $n$-qubit quantum unitaries $U$ and Hamiltonians $H$, given query access to $U$ or the unitary evolution of $H$, that may be dense in the Pauli basis but still admit concise Clifford decompositions. Specifically, we consider unitaries (or Hamiltonians) of the form $U = \sum_i \alpha_i C_i$ over Cliffords $C_i$ with bounded Clifford extent $\sum_i |\alpha_i|$. To extract this Clifford structure, we introduce an agnostic tomography protocol for Clifford unitaries that given query access to an unknown unitary $U$ with optimal Clifford fidelity $\textsf{opt}$, outputs a Clifford unitary witnessing fidelity $\geq \textsf{opt} - \varepsilon$ for some error $\varepsilon > 0$, in time $\textsf{poly}(n,(1/\varepsilon)^{\log(1/\varepsilon)})$. We then apply this protocol to obtain tomography protocols for unitaries and Hamiltonians that have bounded Clifford extent. This extends learnability of Hamiltonians from those with sparse Pauli decompositions to those that are dense (i.e., has sparsity $\Omega(2^n)$) in the Pauli basis but are Clifford structured.
\end{abstract}

\setcounter{tocdepth}{2}
{\small \tableofcontents}

\newpage
\section{Introduction}
Learning the dynamics of an arbitrary $n$-qubit quantum system is an intrinsically high-dimensional task, requiring resources that scale exponentially with $n$. This complexity can be dramatically reduced, however, when the unknown dynamics are known to possess some structure. Several natural classes of quantum unitaries are known to admit efficient tomography algorithms, including Clifford unitaries~\cite{low2009learning,lai2022learning}, Clifford circuits containing few non-Clifford gates~\cite{lai2022learning}, diagonal unitaries from low levels of the Clifford hierarchy~\cite{abdy2023phase}, quantum juntas~\cite{chen2023juntas,bao2023testing}, low-degree unitaries~\cite{arunachalam2024learning}, Pauli-sparse unitaries~\cite{grewal2025query,honjani2026query}, and much more~\cite{huang2024shallow,zhao2024bounded,iyer2025mildly,fanizza2025efficient}. Similarly, one can efficiently learn broad families of Hamiltonians given access to unitary evolution when the Hamiltonian is local or sparse in the Pauli basis \cite{huang2023HL,dutkiewicz2024advantage,caro2024learn,bakshi2024structure,ma2024learning,castaneda2025hamiltonian,zhao2025learning,arunachalam2025testing,sinha2025improved,hu2025ansatz,abbas2025nearly}. These results illustrate a recurring theme in quantum learning theory: although general quantum dynamics are prohibitively complicated, dynamics governed by structured classes and, particularly, concise descriptions over the Pauli operators, may be efficiently learnable.

In practice, the unknown process may not lie exactly in a specified structured class. Noise, implementation errors, or unmodeled effects may cause the process to deviate from the class while still retaining significant correlation with the class, or while admitting a concise decomposition in terms of its members. This motivates the broader question of whether process tomography algorithms can exploit such approximate structure. Clifford unitaries provide a canonical setting for studying this question as they are both efficiently classically simulable~\cite{gottesman1998heisenberg} and learnable. Moreover, \cite{bravyi2016improved} proposed a classical simulation algorithm for quantum circuits with few non-Clifford gates, which are expressible as concise sums over Cliffords, showcasing that this is a model class relative to which the complexity of more general quantum processes can be measured. Thus, understanding whether Clifford structure can be detected and extracted even when the unknown unitary is not itself Clifford is a natural and important direction. Recent work \cite{Gross2017SchurWeylDF,hinsche2025clifford} has shown that we can efficiently test if an unknown unitary $U$ is close to or far from the Clifford group. However, testing does not yield the closest Clifford to $U$, motivating the following question:
\begin{quote}
\centering \emph{Given query access to an unknown $n$-qubit unitary, can we learn the Clifford unitary that is closest to it efficiently?}
\end{quote}
\vspace{-1mm}
This is an instance of \textit{agnostic tomography} of quantum processes: the goal is to output a member of a prescribed model class whose fidelity with the unknown process is nearly optimal. Results along this thread have so far considered model classes with Pauli structure, shallow circuits, and circuits with bounded gate complexity~\cite{wadhwa2025apt,dong2025linear} or have given an \emph{improper} agnostic algorithm for Clifford unitaries where the output is not a Clifford and need not even be unitary~\cite{wadhwa2025apt}. A proper agnostic algorithm for Cliffords may be obtained using~\cite{buadescu2021qda} with polynomial query complexity but may require super-exponential time complexity. Thus, computationally efficient or even sub-exponential proper agnostic tomography of Clifford unitaries remains open.

Beyond recovering a single closest Clifford and motivated by unitaries which can be simulated classically~\cite{bravyi2016improved}, we also consider tomography of unitaries and even Hamiltonians which admit a low-complexity decomposition $U = \sum_{j=1}^k \alpha_j C_j$
over Cliffords $\{C_j\}$ with bounded $\sum_i |\alpha_i|$, often referred to as Clifford extent. This leads to our second core question:
\vspace{-1mm}
\begin{quote}
\centering \emph{Can we efficiently learn a quantum unitary or Hamiltonian that admits a low-complexity decomposition over Clifford unitaries?}
\end{quote}
\vspace{-2mm}
The analogous tomography question of quantum states with bounded extent was only recently investigated~\cite{arunachalam2026tomography} and the stated problem for unitaries and Hamiltonians remains unexplored.

\subsection{Main results}
To describe our main results, we will require a few definitions. Let $\Cliff(n)$ be the class of Clifford unitaries acting on $n$-qubits. Let $\sket{U} := (U \otimes \id) \ket{\EPR_n}$ denote the Choi state of a unitary $U$ obtained by applying $U$ to half of the qubits of $n$-$\EPR$ pairs. For a given $n$-qubit quantum unitary $U$, let
$$
\calF_{\Cliff(n)}(U) := \max_{V \in \Cliff(n)} 2^{-2n} | \Tr(U^\dagger V) |^2 = \max_{V \in \Cliff(n)} | \la \! \la U | V \ra \! \ra|^2,
$$
be the Clifford fidelity of $U$ i.e., the maximal fidelity of $U$ with any Clifford unitary, which is equivalently the maximal fidelity of $\sket{U}$ with the Choi state of any Clifford unitary. The problem of agnostic tomography of Clifford unitaries $\Cliff(n)$ is then to output $V \in \Cliff(n)$ given query access to $U$ such that
$$
2^{-2n} \left| \Tr(U^\dagger V)\right|^2  \geq \calF_{\Cliff(n)}(U) - \varepsilon \quad \text{ or } \quad | \la \! \la U | V \ra \! \ra|^2 \geq \calF_{\Cliff(n)}(U) - \varepsilon,
$$
where $\varepsilon \in (0,1)$ is the desired accuracy. We will call an agnostic learner \emph{weak} if the fidelity guarantee of the output state is $\poly(\calF_{\Cliff(n)}(U))$ and refer to the above definition as a \emph{strong} agnostic learner (see~\Cref{sec:agnostic-learning}).  We call an agnostic tomography protocol \emph{improper} if it outputs a state outside the class $\Cliff(n)$ and proper otherwise. 

\paragraph{Agnostic tomography of Cliffords.} Our first main result is a proper quasipolynomial-time agnostic tomography protocol of Clifford unitaries, which we formally state below.
\begin{restatable}{theorem}{agnosticlearnerCliffords}
\label{thm:agn_learner_cliffords}
Let $\varepsilon,\delta \in (0,1)$. Suppose $U$ is an unknown $n$-qubit unitary with (unknown) optimal Clifford fidelity of $\opt$ ($:= \calF_{\Cliff(n)}(U)$). Then, there is an algorithm that given query access to $U$ and with probability $\geq 1-\delta$, outputs a Clifford unitary $V$ such that
$$
| \la \! \la U | V \ra \! \ra|^2 \geq \opt - \varepsilon.
$$
The query and time complexity of this algorithm is 
$\poly(n, (1/\varepsilon)^{\log(1/\varepsilon)}, \log (1/\delta))$.
\end{restatable}
This resolves an open question of proper agnostic tomography of Cliffords~\cite{wadhwa2025apt}. Previous work only gave an efficient \emph{improper} agnostic algorithm for Cliffords whose output may not be a Clifford nor a unitary \cite{wadhwa2025apt}. Our agnostic tomography algorithm crucially uses the agnostic tomography algorithm of stabilizer states in \cite{chen2025stabilizer} as a subroutine and inherits its complexity, but instead of directly outputting the $2n$-qubit stabilizer state obtained from agnostic tomography of the Choi state $\sket{U}$, which may not be a valid Choi state of an $n$-qubit Clifford, we give a protocol that outputs a valid $n$-qubit Clifford witnessing near-optimal fidelity. This will be described in Section~\ref{sec:tech_overview}. The agnostic tomography protocol also implies a quasipolynomial-time tolerant tester of Clifford unitaries up to $1/\poly(n)$ additive gap.

We expect our proper agnostic tomography algorithm to have applications in circuit verification and error mitigation. Properness is relevant here because the output can then be directly compiled, simulated, and compared with an intended circuit. For circuit verification, our algorithm can be viewed as learning the best Clifford explanation of an implemented circuit. If a device is intended to implement a Clifford circuit, the learned Clifford identifies the closest ideal Clifford behavior and can reveal coherent Clifford-frame or compilation errors, complementing fidelity-estimation and randomized benchmarking methods~\cite{flammia2011direct,dasilva2011practical,magesan2011scalable}. The learned Clifford also provides a classically simulable surrogate which can be utilized for error mitigation in the spirit of Clifford-data regression and probabilistic error-cancellation based approaches~\cite{temme2017error,endo2018practical,czarnik2021error,lowe2021unified}.

\paragraph{Tomography for dynamics with bounded extent.} Our second main result is tomography algorithms for $n$-qubit Clifford-structured unitaries and Hamiltonians. Specifically, we consider unitaries (and Hamiltonians) that have Clifford decompositions of the form $U = \sum_{i=1}^M \alpha_i C_i$ over Clifford unitaries $\{C_i\}$ for some $M \in \mathbb{N}$ such that $\sum_i |\alpha_i|$, which is often called the Clifford extent~\cite{bravyi2016improved}, is bounded. The minimal parameter $M$ over all possible Clifford decompositions of $U$ is called the Clifford rank. We work with the \emph{normalized Frobenius norm} defined as $\norm{A}_{\overline 2} := \sqrt{\Tr(A^\dagger A)/2^n}$. Formally, we show the following.
\begin{restatable}{theorem}{learnlowcliffextent}
\label{thm:learn_unitaries_low_cliff_extent}
Let $\varepsilon,\delta \in (0,1)$ and $\xi \geq 1$. Suppose $U$ is an unknown $n$-qubit unitary and $H$ is an unknown $n$-qubit Hamiltonian, both with Clifford extent $\leq \xi$. Then, there exist the following algorithms to learn $U$ and $H$.
\begin{enumerate}[$(i)$]
\item (See~\Cref{cor:learning_low_clifford_extent}) There is an algorithm that given query access to $U$ and with probability $\geq 1-\delta$, outputs an operator $V$ with Clifford rank $O(\xi^2/\varepsilon^2)$ such that
$$
\min_{\theta \in [0,2\pi)} \norm{V - e^{i\theta}U}_{\overline 2} \leq \varepsilon.
$$
The query and time complexity of this algorithm is $\poly(n, (\xi/\varepsilon)^{\log(\xi/\varepsilon)}, \log(1/\delta))$.
\item (See~\cref{thm:HL_low_clifford_extent}) There is an algorithm that given query access to $\exp(-iHt)$ and with probability $\geq 1-\delta$, outputs a Hamiltonian $\widehat{H}$ with Clifford rank $O(\xi^4/\varepsilon^4)$ such that
$$
\norm{\widehat{H} - H}_{\overline 2} \leq \varepsilon.
$$
The algorithm uses $\poly(n, (\xi/\varepsilon)^{\log(\xi/\varepsilon)}, \log(1/\delta))$ query complexity, time complexity and time evolution.
\end{enumerate}
\end{restatable}

This identifies a new learnable regime for quantum dynamics: processes that may be highly non-sparse in the Pauli basis, but are nevertheless sparse or low-complexity when expressed over Clifford unitaries. This is qualitatively different from much of the existing literature on Hamiltonian learning, where the relevant structure is locality and sparsity in the Pauli basis \cite{yu2023robust,huang2023HL,dutkiewicz2024advantage,caro2024learn,bakshi2024structure,ma2024learning,castaneda2025hamiltonian,zhao2025learning,arunachalam2025testing,sinha2025improved,hu2025ansatz,abbas2025nearly}. For example, even a single Clifford unitary can have exponentially many nonzero Pauli coefficients, as witnessed by $(H^{\otimes n}=2^{-n/2}\sum_{S\subseteq[n]}X_S Z_{\overline S})$, while having Clifford extent one. Thus, bounded Clifford extent captures a form of structure that is not amenable to Pauli-analytic methods. Our algorithms therefore extend the scope of efficient process tomography from dynamics with concise Pauli descriptions to dynamics with concise Clifford decompositions, matching quantum circuits known to be classically simulable~\cite{bravyi2016improved,bravyi2019simulation}. Conceptually, this provides a process-level analogue of recent tomography results for quantum states with bounded extent~\cite{arunachalam2026tomography}, and shows that low-complexity linear combinations of efficiently simulable quantum processes (Cliffords) are themselves learnable.

\subsection{Technical overview}\label{sec:tech_overview}

\subsubsection{Agnostic tomography of Clifford unitaries}

\paragraph{\emph{How much progress does stabilizer bootstrapping make?}}

Our starting point is the stabilizer bootstrapping algorithm of \cite{chen2025stabilizer}, which outputs a list of stabilizer states which have high fidelity with an unknown input state. A natural approach to learning the closest Clifford to an unknown unitary $U$ is to prepare its Choi state $\sket{U}$, apply the stabilizer bootstrapping algorithm to learn a stabilizer state $\ket\phi$ with high fidelity to $\sket{U}$, and then invert $\ket\phi$ under the Choi map to obtain the (approximately) closest Clifford. The problem with this approach is that $\ket\phi$ is not guaranteed to be a Choi state of a Clifford, i.e. a maximally entangled stabilizer (the matrix obtained by applying the inverse Choi map to $\ket\phi$ may not be unitary). Indeed, while stabilizer bootstrapping is guaranteed to output the closest stabilizer state to $\sket{U}$, there can be unitaries for which the closest stabilizer state to $\sket{U}$ is not itself maximally entangled.

The next natural approach would be to search in the neighborhood of the output state $\ket\phi$ for maximally entangled stabilizer states with high fidelity to $\sket{U}$. In particular, if $U$ has Clifford fidelity $\opt$, then any stabilizer state $\ket\phi$ with fidelity at least $\opt$ must have entanglement entropy at least $n- O(\log(1/\opt))$ across the Choi cut~\cite{hinsche2025clifford}. Furthermore, any stabilizer state with entanglement entropy $k$ can be written as a superposition of two stabilizer states with entanglement entropy $k \pm 1$, and therefore $\ket\phi$ can be written as a sum over at most $2^{O(\log1/\opt)}$ maximally entangled stabilizer states. By an averaging argument, there is a nearby stabilizer state $\ket{\phi'}$ with entanglement entropy $n$, and ``high'' fidelity to $\sket{U}$. Furthermore, finding such a $\ket{\phi'}$ is efficient. However, the averaging argument incurs a loss in fidelity of $\opt$, which only gives a weak agnostic learner whose final fidelity is quadratic in the optimum (See \Cref{sec:wal_appendix} for details).

\paragraph{Our approach.}
We overcome this problem by applying a more involved analysis to the stabilizer bootstrapping procedure. This analysis follows a similar approach to \cite{briet2026near} (Lemma 4.1). Note that the above weak agnostic learner finds a stabilizer state with fidelity at least $\opt$ and entanglement entropy $n-k$ for some $k$, and then ``rounds'' to a nearby maximally entangled stabilizer state, incurring a loss in fidelity of at most $2^k$.
Our main technical insight is the observation that whenever stabilizer bootstrapping is overwhelmingly likely to output a state with entanglement entropy $n-k$, that state must have fidelity $\approx 2^k \opt$. Therefore, ``rounding'' back to a maximally entangled stabilizer state and incurring a loss in fidelity of $2^k$ yields a state with fidelity $\approx \dfrac{2^k \opt }{2^k} = \opt$ to $\sket{U}$. 

Concretely, we analyze the \emph{stabilizer-neighbor graph}. The vertices of this graph are stabilizer states, with an edge between two stabilizer states whenever their fidelity is $1/2$. Importantly, for any stabilizer state with fidelity at least $\opt$ which approximately maximizes fidelity among all its neighbors (these states are called $\gamma$-approximate local maximizers), the stabilizer bootstrapping algorithm is guaranteed to output this state with nonnegligible probability. Now, suppose we start from a Clifford Choi state $\sket{C}$ having fidelity $\opt$ with the unknown input $\sket{U}$. If $\sket{C}$ is not a $\gamma$-approximate local maximizer, then there is a neighboring stabilizer state whose fidelity with $\sket{U}$ is larger by a factor of at least $1/\gamma$. We can continue following such improving edges until we reach a $\gamma$-approximate local maximizer $\ket{\phi}$. Since each step increases the fidelity by a factor of at least $1/\gamma$, and fidelity is at most one, this path has length $t=O(\log(1/\opt))$. The list-decoding guarantee of stabilizer bootstrapping allows us to find $\ket\phi$ in quasipolynomial time.

The crucial observation is that the neighbor graph also controls how far $\ket{\phi}$ can be from a Clifford Choi state. In particular, we show that moving across a single edge can change the entanglement entropy by at most one. Since $\sket{C}$ is maximally entangled and $\ket{\phi}$ is at distance at most $t$ from $\sket{C}$ in the neighbor graph, $\ket{\phi}$ can be written as a superposition of at most $t$ maximally-entangled stabilizer states. Using the normal form of bipartite stabilizer states~\cite{fattal2004entanglement}, we can therefore round $\ket{\phi}$ back to a maximally entangled stabilizer state by searching over at most $2^t$ possibilities. By an averaging argument, one of these Clifford Choi states is guaranteed to lose at most a factor $2^t$ in fidelity. On the other hand, the path to $\ket{\phi}$ increased the fidelity by a factor of at least $\gamma^{-t}$. Combining these two effects, we obtain a Clifford Choi state with fidelity at least
\[
    \frac{\opt}{(2\gamma)^t}.
\]
We then choose $\gamma$ to be only slightly larger than $1/2$. Since $t=O(\log(1/\opt))$, we can choose $\gamma$ so that $(2\gamma)^t$ is sufficiently close to one, giving final fidelity at least $\opt-\varepsilon$. Overall, the path through the neighbor graph simultaneously bounds the loss of entanglement and guarantees a corresponding increase in fidelity, which allows us to obtain a strong agnostic tomography algorithm for Clifford unitaries.

\subsubsection{Tomography for unitaries and Hamiltonians with bounded Clifford extent}
\paragraph{Tomography for unitaries with bounded extent.} 
Our first application is an algorithm that takes copies of an unknown $n$-qubit unitary $U$ and outputs a list of $k=O(1/\varepsilon^2)$ many $n$-qubit Clifford unitaries $\{V_i\}_{i \in [k]}$ such that the Choi state $\sket{U}$ can be expressed as a compact superposition of $\{\sket{V_i}\}_{i \in [k]}$ and a residual $2n$-qubit state that is promised to have low fidelity ($\leq \varepsilon$) with the Choi state of any $n$-qubit Clifford. This is obtained as a direct consequence of applying our agnostic tomography protocol of Clifford unitaries along with the algorithmic decomposition result of~\cite{arunachalam2026tomography}. The resulting protocol uses time and sample complexity that grows polynomial in the number of qubits and inverse quasipolynomial in the approximation error $\varepsilon$. This partially answers the question of \cite{bu2025quantum} that conjectured the existence of such Clifford decompositions (but directly in the unitary form). When the unknown unitary is promised to have bounded Clifford extent, we show that the linear operator $\sum_{i \in [k]} V_i$ accomplishes the task of tomography.

\paragraph{Tomography for Hamiltonians with bounded extent.}
Our second application is tomography for $n$-qubit Hamiltonians $H$ under the promise that $\Tr(H)=0$ (i.e., does not have an identity term) and $H$ has bounded Clifford extent. Our main observation is that if $H$ is a Hamiltonian with Clifford extent $\leq \xi$, then the unitary evolution $U(t) = \exp(-iHt)$ has Clifford extent $\leq e^{t \xi}$. This bound can be made small by choosing a suitable $t \leq 1/\xi$. We can then apply our tomography algorithm for unitaries with bounded extent from above to learn an approximation $\widehat{U}(t)$. Noting that the Taylor expansion of $U(t) = \id - iHt + O(t^2)$, we can then obtain an estimate $\widehat{H}$ of $H$ by setting $\widehat{H} = (\id - \widehat{U}(t))/(it)$. The main technical work involved then is to show that $\widehat{H}$ is close to the true $H$ under the normalized Frobenius norm by analyzing the truncation error from the Taylor series. The resulting algorithm has time complexity $\poly(n,(\xi/\varepsilon)^{\log(\xi/\varepsilon)})$.

The procedure above relies on a Hamiltonian $H$ having small Clifford extent. We observe that there are instances where the stabilizer extent of $\sket{e^{-iH t}}$ is much smaller than the Clifford extent of $H$ itself. We therefore formulate an alternative Hamiltonian learning algorithm based on learning the stabilizer decomposition of $\sket{e^{- i Ht}}$. We give a concrete example of the following form 
$$
H = \sum_{j,k} h_{j,k}\ket{s_j} \bra{s_k}
$$
where $\ket{s_j}$ are stabilizer states and the coefficients $h_{jk}$ for a Hermitian matrix ${\bf h} \in \mathbb{C}^{R \times R}$. We show that while the time and sample complexity of the former algorithm can be as large as $\poly (n (R/\varepsilon)^{\log (R/\varepsilon)})$, if $R \ll 2^{n/3}$, the corresponding complexity for the alternative algorithm can be made $\poly (n(g^*h^*/\varepsilon)^{\log ((g^* h^*/\varepsilon))})$, where $h^*$ is the largest eigenvalue of ${\bf h}$ and $g^*$ is the largest eigenvalue of the Gram matrix of $\{\ket{s_j}\}_j$ which in principle can be made independent of $R$. 

The example above raises a natural question: what are the fundamental relationships between Clifford extent of an operator and the stabilizer extent of the corresponding Choi state. The above example shows they can be in general far apart for the Hamiltonian described above. We prove the existence of unitary operators for which the two parameters are exponentially separated (\Cref{app:xi-separations}). 

\paragraph{Examples of dense Hamiltonians with bounded extent.} Current Hamiltonian tomography algorithms assume that the Hamiltonian is sparse in the Pauli basis \cite{arunachalam2025testing, ma2024learning}. On the other hand, we have given an algorithm that learns Hamiltonians with bounded Clifford extent in quasipolynomial time. An immediate question is then what classes of Hamiltonians can now be learned by leveraging Clifford structure which were previously not amenable to algorithms that exploited Pauli structure? 

One such class of Hamiltonians is the sum of exchange interactions (which swaps subsets of qubits with size $k$). Each exchange interaction is exponentially dense in $k$ when written as a sum of Paulis, but is itself a single Clifford operator. One can also design efficient algorithms for families of Hamiltonians that can be written as sum of a few stabilizer projectors. Classes of Hamiltonians that are non-trivially sparser in this setting than the Pauli basis are Hamiltonians formed from sums of mutually commuting Pauli subgroups, and Hamiltonians that are sparse in the computational basis.

\subsection{Open questions}
\label{sec:open_questions}
Our work opens up several interesting directions for future work.
\begin{enumerate}
    \item \emph{Improved algorithms and lower bounds:} The agnostic protocols for Clifford unitaries proposed in this work utilize agnostic tomography protocols of stabilizer states \cite{chen2025stabilizer} as a subroutine. The query access to the unknown unitary is primarily used to create the corresponding Choi state. One could then imagine that perhaps an agnostic tomography protocol with improved time complexity could be obtained by utilizing this query access more elaborately. Moreover, as we use the Choi state, we require $n$ ancilla qubits. Another question would be to investigate what can be accomplished using limited quantum memory as has been shown recently for testing Clifford unitaries~\cite{hinsche2025clifford}.  Finally, the output of our tomography protocol of unitaries with bounded Clifford extent is a linear operator, $V$, that is not necessarily unitary. An immediate extension would be to obtain a proper protocol.
    \item \emph{Tomography of other classes:} For what other classes of quantum unitaries can we develop efficient strong agnostic tomography algorithms? One natural direction is to extend our result for the classically simulable Clifford unitaries to richer circuit classes, such as unitaries implemented by Clifford circuits with few $T$ gates or even fermionic Gaussian unitaries. This in turn should yield tomography algorithm for unitaries that would admit decompositions over these classes and have bounded extent with respect to them.
    \item \emph{Long-time Hamiltonian tomography:} The algorithm developed in this work for tomography of Hamiltonians with bounded Clifford extent relies on access to arbitrarily short-time evolutions. Since such control is difficult to achieve in practice, an important direction for future work is to eliminate this requirement as has been recently investigate for learning Pauli-sparse Hamiltonians \cite{shin2026heisenberg,de2026learning}.
    \item What are the fundamental relationships between Clifford complexity of a unitary $U$ and the stabilizer complexity of its corresponding Choi state $\sket{U}$? See \cref{def:extent/rank} for the definitions. In \Cref{app:xi-separations} we showed that the corresponding Clifford and stabilizer extents can be exponentially separated. An immediate question is how this separation depends on approximation error if we consider approximate notions of extent (say $\xi^{(\delta)}_{\Cliff}$ and $\xi^{(\delta)}_{\Stab}$). We expect the result to be robust if we consider operator norm for the distance between unitaries. The argument however may not be as robust if we consider metrics such as normalized Frobenius norm. 
    Can we show the inverse connection, i.e.,
    $$
    \xi^{(\delta)}_{\Cliff} (U)\leq \left(\xi^{(\delta)}_{\Stab} (\sket{U})\right)^C
    $$
    for some constants $C, \delta$ for natural definitions for approximation? 
    We can ask the same questions about the connection between Clifford rank of $U$ and the stabilizer rank of $\sket{U}$. We believe the rank quantities are also exponentially separated. Consider for example the long range controlled $Z$ operator $C^nZ = \sum_{x \in\{0,1\}^n} (-1)^{x_1 \ldots x_n} \ket{x}\bra{x}$. We can show $C^n Z = I - 2 \ket{1^n}\bra{1^n}$ therefore the stabilizer rank of $\sket{C^n Z}$ is $2$. However to decompose $C^n Z$ into Clifford unitaries one will likely need to decompose the high degree AND, i.e., $(-1)^{x_1 \ldots x_n}$ phase, into quadratic phases. We know that in terms of linear phases we need exponentially large decompositions. For quadratic phases this decomposition is conjectured to require an exponential number of terms (this is known as the quadratic uncertainty principle \cite{filmus2014real}). Can we show an exponential separation between Clifford and stabilizer ranks assuming the quadratic uncertainty principle? We leave this as an open question.
\end{enumerate}

\subsubsection{Organization}
The rest of the paper is organized as follows. In \cref{sec:prelims}, we review some basic notation, and define terminology allowing us to state the problem of agnostic tomography of Clifford unitaries. In \cref{sec:agnostic-learning-cliffords}, we give our main contribution which is a strong agnostic tomography protocol for Clifford unitaries. In doing so, we actually give a more general scheme for agnostic tomography of $2n$-qubit stabilizer states with entanglement entropy $k$ across a fixed $n$-qubit bipartition. Finally in \cref{sec:applications}, we give some applications for our strong agnostic tomography protocol, namely tomography of unitaries and Hamiltonians with bounded Clifford extent. 

\paragraph{Acknowledgements.} The authors thank Srinivasan Arunachalam, who was involved in the initial stages of this project, for multiple discussions on agnostic learning. AD thanks Jop Bri\"et for discussions on agnostic learning of Cliffords. We acknowledge using ChatGPT $\leq 5.5$ Plus for combinatorial bounds regularly, and for suggesting the proof of \Cref{thm:extent-separation}, which the authors verified and then rewrote in \Cref{app:xi-separations}. AD acknowledges the idea for obtaining the phase in Theorem~\ref{thm:HL_low_clifford_extent} was obtained from ChatGPT 5.6 Plus. Specifically, a hand-written proof of Theorem~\ref{thm:HL_low_clifford_extent} was given to ChatGPT assuming Corollary~\ref{cor:learning_low_clifford_extent} outputs a unitary with the correct global phase, upon which it gave the suggestion. This was incorporated, verified and completely hand-written by the authors. SM and DJ are grateful to the National Science Foundation (NSF CCF-2013062) for supporting this
project.

\newpage 

\section{Preliminaries}
\label{sec:prelims}
\subsection{Basic definitions and notation}
\paragraph{Basic Notation.} 
Throughout the paper we use notation $[n] = \{1,\ldots, n\}$ and $[n]_0 = \{0, 1,\ldots, n\}$. $\widetilde{O}$-notation is similar to $O$, but ignores logarithmic factors. All logarithms are base $2$, unless stated otherwise.

\paragraph{Fidelity between quantum states.}
The fidelity between density matrices $\rho$ and $\sigma$ is defined as $F(\rho, \sigma) = \Tr (|\sqrt{\rho} \sqrt{\sigma}|)^2$. For pure states $\ket{\psi}$ and $\ket{\phi}$ the fidelity is $F (\psi, \phi) = \left|\braket{\psi|\phi}\right|^2$. 

\paragraph{Choi states.} For a matrix $M$ acting on $n$ qubits the corresponding Choi state vector is defined as
$$
\sket{M} := (M \otimes \id) \ket{\mathsf{EPR}_n}.
$$
This defines an isomorphism between matrices $M$ and  quantum states $\sket{M}$ using
$$
\sket{M} = \frac{1}{\sqrt{2^n}}\sum_{x,y} M_{x,y} \ket{x
} \ket{y} \leftrightarrow M = \sum_{x,y} M_{x,y} \ket{x
} \bra{y}.
$$
\paragraph{Entanglement entropy.} For a quantum state $\ket{\psi}_{AB}$ defined across two subsystems $A$ and $B$ the entanglement entropy across this cut is defined as 
$$
E_A(\ket{\psi}_{AB}) = - \Tr (\rho_A \log_2 \rho_A) = -\Tr (\rho_B \log_2 \rho_B),
$$
where $\rho_A = \Tr_B(\ket{\psi}_{AB}\bra{\psi})$ is the reduced density matrix on part $A$; $\rho_B$ is defined similarly.
 
\paragraph{Norms.} We will primarily work with linear operators from $\mathbb{C}^N$ to $\mathbb{C}^N$ where $N=2^n$ and $n$ represents the dimension of the quantum system being considered. We will denote the corresponding space as $\calM_{N,N}$ and set of unitaries in $\calM_{N,N}$ as $\mathcal{U}_N$. We will denote $\norm{A}_{\op}$ as maximum singular value of $A$ (or equivalently the induced matrix norm) i.e., $\norm{A}_{\op} := \sigma_{\max}(A) = \sup_{x : \norm{x}_2 = 1} \norm{Ax}$ where $\sigma_\max(A)$ represents the maximum singular value of $A$. We will work with the $p$-Schatten norms of the matrices defined as
$$
\|A\|_p = \left(\Tr (|A|^p)\right)^{1/p}.
$$
For $p=2$, this yield the Frobenius norm of $A$ i.e., $\norm{A}_2 = \sqrt{\Tr(A^\dagger A)}$. We will also work with the \emph{normalized Frobenius norm} defined as
\begin{equation}\label{def:normalized_frob_norm}
\norm{A}_{\overline{2}} := \sqrt{\Tr(A^\dagger A)}/\sqrt{N}.
\end{equation}
represented by a subscript $\overline{2}$. We will use the following metric to compare the distance between unitary matrices (or linear operators in the same space).
\begin{definition}\label{def:dist_linear_ops}
Given $A,B \in \calM_{N,N}$, we define
$$
\dist(A,B) := \min_{\theta \in [0,2\pi)} \frac{1}{\sqrt{2N}} \norm{e^{i\theta} A - B}_2.
$$
\end{definition}
Note that $\dist(A,B)^2 = 1 - |\la\!\la A | B \ra\!\ra|$ when $A,B \in \mathcal{U}_N$. The above definition is useful when we learn unitaries only up to a global phase. We will need the following fact regarding the normalized trace.
\begin{fact}\label{fact:normalized_trace}
Let $A,B \in \calM_{N,N}$ and define $\mu(A) := \Tr(A)/N$. Then, $A \mapsto \mu(A)$ is $1$-Lipschitz with respect to the normalized Frobenius norm:
$$
|\mu(A) - \mu(B)| \leq \norm{A - B}_{\overline 2}.
$$
\end{fact}
\begin{proof}
Let $D = A - B$. Using Cauchy-Schwarz, we have $|\Tr(D)| = |\Tr(\id^\dagger D)| \leq \norm{\id}_2 \norm{D}_2 \leq \sqrt{N} \norm{D}_2$. This implies $|\Tr(D)|/N \leq \norm{D}_2/\sqrt{N} = \norm{D}_{\overline 2}$. Substituting back for $D$ gives us the desired result.
\end{proof}

\subsection{Complexity measures based on Clifford and stabilizer structures}
We first start with basic definitions of stabilizer states and Clifford unitaries.
\paragraph{Stabilizer states.} The Pauli operators over qubits are defined as
$$
I = \begin{pmatrix}
    1 & 0 \\ 0 & 1
\end{pmatrix}, \quad X = \begin{pmatrix}
    0 & 1 \\ 1 & 0
\end{pmatrix}, \quad
Y = \begin{pmatrix}
    0 & -i \\ i & 0
\end{pmatrix}, \quad
Z = \begin{pmatrix}
    1 & 0 \\ 0 & -1
\end{pmatrix}.
$$
The Pauli group over $n$ qubits is defined as 
$$
\mathcal{P}_n = \{c P_1 \otimes \ldots \otimes P_n : P_j \in \{I, X, Y, Z\}, c \in \{\pm 1, \pm i\} \}.
$$
A quantum state is called a stabilizer state if it is stabilized by an abelian subgroup of the Pauli group of size $2^n$ not containing $-I$. We denote the set of stabilizer states over $n$ qubits as $\Stab (n)$.

Additionally, we will require the ability to efficiently compute the fidelity between an unknown state $\rho$ and a set of stabilizer states~\cite{huang2020predicting}.
\begin{lemma}\label{lem:shadows_stabilizers}
Given copies of an unknown $n$-qubit quantum state $\rho$ and classical descriptions of $M$ stabilizer states $\ket{\phi_1}, \dots, \ket{\phi_M}$, there is an algorithm that, with probability at least $1-\delta$, estimates $\bra{\phi_i}\rho\ket{\phi_i}$ to additive error at most $\varepsilon$ for all~$i$. The sample complexity is $O(\frac{1}{\varepsilon^2}\log \frac{M}{\delta})$ and the time complexity is $O(\frac{Mn^2}{\varepsilon^2}\log \frac{M}{\delta})$.
\end{lemma}

\paragraph{Clifford circuits.} The Clifford group is generated by the basic gates 
$$
H = \frac{1}{\sqrt{2}} \begin{pmatrix}
    1 & 1 \\ 1 & -1
\end{pmatrix}, \quad S = \begin{pmatrix}
    1 & 0 \\ 0 & i
\end{pmatrix}, \quad CNOT = \begin{pmatrix}
    1 & 0 & 0 & 0 \\ 0 & 1 & 0 & 0\\ 0 & 0 & 0 & 1 \\ 0 & 0 & 1 & 0
\end{pmatrix}.
$$
It corresponds to the normalizer of the Pauli group, meaning Clifford unitaries map Pauli group elements to Pauli group elements under conjugation. Consequently, Clifford unitaries map stabilizer states to stabilizer states. Due to this property, starting from a stabilizer state the action of a Clifford circuit can be efficiently simulated on a classical computer~\cite{gottesman1998heisenberg}. We denote the Clifford group over $n$ qubits with $\Cliff (n)$.

\subsubsection{Measures of stabilizer complexity}
\paragraph{Stabilizer and Clifford fidelity.}
Let $\mathcal{S}$ be a family of quantum states (which can be pure states or density matrices). For any $n$-qubit state $\rho$ and set of $n$-qubit states $\mathcal{S}$, we define the $\mathcal{S}$-fidelity of $\rho$ to be 
\begin{equation}
 \mathcal{F}_{\mathcal{S}} (\rho) := \max_{s \in \mathcal{S}}F(s, \rho).
\end{equation}

When $\mathcal{S}$ is the set of $n$-qubit stabilizer states we refer to this as the \textit{stabilizer fidelity}. To define the analogous notion for unitaries we use the Choi map, which maps unitaries to states. In particular, for a unitary matrix $U$ consider the corresponding Choi state on $2n$ qubits,
\begin{equation}\label{def:choi_state_U}
    \sket{U} = (U \otimes \id) \ket{\EPR_n} = (U \otimes \id) \frac{1}{\sqrt{2^n}}\sum_{x \in F_2^{n}}\ket{x}\ket{x}.
\end{equation} 
Now, letting $\Cliff (n)$ be the set of $n$-qubit Clifford unitaries, for any $n$-qubit unitary $U$ we define the Clifford fidelity as 

\begin{equation}
    \mathcal{F}_{\Cliff(n)} (U):=\max_{C \in \Cliff(n)}F(\sket{U}, \sket{C}).
\end{equation}

\paragraph{Stabilizer/Clifford rank and extent.} Fidelity measures correlation between an object and a set. We can measure complexity of an object also in terms of the sparsity of its decomposition in terms of elements of a set. 
\begin{definition}[Stabilizer and Clifford rank/extent]\label{def:extent/rank}
    For a quantum state $\ket{\psi} \in (\mathbb{C}^2)^{\otimes n}$
    \begin{itemize}
        \item Its stabilizer rank is defined as 
        \begin{equation}
        \chi_{\Stab}(\ket{\psi}) := \min \left \{ \|c\|_{0} \hspace{1mm} : \hspace{1mm}  \ket {\psi} = \sum_{s \in \Stab (n)}c_s \ket{s}  \right \}.
    \end{equation}
        \item Its stabilizer extent is defined as
    \begin{equation}
    \xi_{\Stab} (\ket{\psi}) := \min \left \{ \|c\|_1 \hspace{1mm}: \hspace{1mm} \ket {\psi} = \sum_{s \in \Stab (n)}c_s \ket{s}\right \}.
    \end{equation}
    \end{itemize}
Similarly for an operator $U$ defined over $n$ qubits 
\begin{itemize}
        \item Its Clifford rank is defined as 
        \begin{equation}
        \chi_{\Cliff}(U) := \min \left \{ \|c\|_{0} \hspace{1mm}:\hspace{1mm} U = \sum_{V \in \Cliff (n)}c_V V\right \}.
    \end{equation}
        \item Its Clifford extent is defined as
    \begin{equation}
        \xi_{\Cliff}(U) := \min \left \{ \|c\|_1 \hspace{1mm}:\hspace{1mm} \ket {\psi} = \sum_{V \in \Cliff (n)}c_V V \right \}.
    \end{equation}
    \end{itemize}
\end{definition}

We note that for an operator $U$
$$
\chi_{\Stab} (\sket{U})\leq \chi_{\Cliff} (U), \quad \xi_{\Stab} (\sket{U})\leq \xi_{\Cliff} (U).
$$
It is interesting to see if these inequalities hold in the reverse, i.e., upper bounding Clifford rank/extent of of a unitary with a polynomial in terms of the stabilizer rank/extent of its Choi state. As a matter of fact we can show (see \Cref{app:xi-separations}) that $\xi_{\Stab} (\sket{U})$ and $\xi_{\Cliff} (U)$ can be exponentially separated. Similar separations might also hold for rank (See the open questions in \cref{sec:open_questions}).

We can also define mixed rank/extent for operators.
\begin{definition}
    For an operator $U$ defined over $n$ qubits, define the mixed rank $\chi_{\mathrm{mixed}} (U)$ of a unitary $U$ as the minimal $\delta$-approximate decomposition in terms of Clifford unitaries, stabilizer projectors and products of the two (e.g., $C_1 \Pi_2 C_3$, where $C_1$ and $C_2$ are Clifford and $\Pi$ is a stabilizer projector).
\end{definition}

From here we can prove the following lemma.
\begin{lemma}
    The mixed rank of an operator $U$ is precisely equal to the stabilzier rank of its Choi state:
    $$
    \chi_{\mathrm{mixed}} (U) = \chi_{\Stab} (\sket{U}).
    $$
\end{lemma}
\begin{proof}
The inequality 
$$
\chi_{\Stab} (\sket{U}) \leq \chi_{\mathrm{mixed}} (U)
$$
is again immediate because the mixed family maps stabilizer states to stabilizer states (up to a proportionality constant). To see  
$$
\chi_{\Stab} (\sket{U}) \geq \chi_{\mathrm{mixed}} (U)
$$
we note that any stabilizer state can be written as product $C \Pi \ket{\EPR_n}$ \cite{fattal2004entanglement}.
\end{proof}

One can also consider definitions of approximate rank/extent by replacing equality between the object and a decomposition with an approximation with respect to a suitable norm (see for instance~\cite{bravyi2016improved}). But for the purpose of this work we mainly focus on exact rank/extent quantities. 

The relationship between stabilizer extent and rank have been investigated in~\cite{mehraban2025improved,mehraban2024quadratic,kalra2026stabilizer}. We will use the following result.  
\begin{theorem}[{\cite[Theorem~31]{kalra2026stabilizer}}]\label{thm:ub_stab_extent_stab_rank_states}
Let $\ket{\psi}$ be an $n$-qubit pure state such that $\ket{\psi} = \sum_{i=1}^k c_i \ket{s_i}$ where $\ket{s_i} \in \Stab$ are linearly independent stabilizer states with corresponding coefficients $c_i \in \mathbb{C}$. Then, we have that 
$$
\xi_\Stab(\ket{\psi}) \leq \sum_{i=1}^k |c_i| \leq O((2 k)^{(2k + 1)/2}).
$$
\end{theorem}
    
\subsection{Useful tomography results}
\subsubsection{Agnostic tomography}
\label{sec:agnostic-learning}
The problem of agnostic tomography of quantum states is as follows: Given copies of an unknown $n$-qubit quantum state $\ket{\psi}$ with unknown optimal fidelity $\geq \tau$ with a specified class of states $\calC$, output a state $\ket{\phi} \in \C$ such that $|\la \phi | \psi \ra|^2 \geq \tau - \varepsilon$ for some error $\varepsilon > 0$. If the agnostic algorithm outputs a hypothesis state that is not in $\calC$ but witnesses the fidelity guarantee, we call the algorithm $\emph{improper}$. Moreover, we call the algorithm a \emph{weak} agnostic tomography protocol if it outputs a state in $\C$ that witnesses fidelity $\tau^C$ for some $C > 1$. We call the agnostic tomography task described above \textit{strong} agnostic tomography to distinguish it from weak agnostic tomography.

\paragraph{Stabilizer states.} We will use the stabilizer bootstrapping algorithm of \cite{chen2025stabilizer} which accomplishes strong agnostic tomography of stabilizer states. Formally, they show the following.
\begin{theorem}[Stabilizer bootstrapping \cite{chen2025stabilizer}]
\label{thm:SB}
Fix $\delta \in (0, 1)$ and $  \varepsilon\leq \tau \in (0,1)$. 
There is an algorithm that, given~copies of $\ket{\psi}$ with $\Fe_{\Stab}({\ket{\psi}}) \geq \tau$, with probability $1-\delta$ outputs a $|\phi\rangle \in \Stab$ with $|\langle \phi | \psi \rangle|^2 \geq \tau - \varepsilon$. The sample~and time complexity of this algorithm is $n\cdot \poly(1/\varepsilon,\log(1/\delta),(1/\tau)^{\log 1/\tau})$ and $\poly(n,1/\varepsilon,\log(1/\delta), (1/\tau)^{\log 1/\tau})$~respectively.
\end{theorem}
Additionally, we will require their \emph{list decoding} result.
\begin{definition}[$\gamma$-approximate local maximizer \cite{chen2025stabilizer}]
Fix an $n$-qubit state $\rho$. For $\gamma > 0$, a stabilizer state $|\phi\rangle \in \mathcal{S}$ is a $\gamma$-approximate local maximizer of fidelity with $\rho$ if
\[
    \langle\phi|\rho|\phi\rangle \geq \gamma \max_{\substack{|\phi'\rangle \in \mathcal{S}, \\ |\langle\phi'|\phi\rangle| = \frac{1}{\sqrt{2}}}} \langle\phi'|\rho|\phi'\rangle.
\]
That is, $|\phi\rangle$ approximately maximizes fidelity over its neighbors $|\phi'\rangle$ in $\mathcal{S}$.
\end{definition}

\begin{theorem}[{\cite[Theorem~6.1]{chen2025stabilizer}}]\label{thm:stabilizer_bootstrapping}
Fix $\tau > 0$ and $1/2 < \gamma \le 1$, and let $\rho$ be an unknown $n$-qubit state. There is an algorithm with the following guarantee. Let $|\phi\rangle$ be any $\gamma$-approximate local maximizer of fidelity with $\rho$, and suppose its fidelity with $\rho$ is at least $\tau$. Given copies of $\rho$, the stabilizer bootstrapping algorithm outputs $|\phi\rangle$ with probability at least $((\gamma - 1/2)\tau)^{O(\log \frac{1}{\tau})}$.
\end{theorem}

\subsubsection{Tomography of bounded extent states}
Given an agnostic tomography protocol of a class $\C$, it was shown in \cite{arunachalam2026tomography} that one can obtain tomography protocols for quantum states with bounded extent defined with respect to $\C$. To formally describe this result, let us first introduce the definition of a \emph{model class}.
\begin{definition}[{\cite[Definition~1.1]{arunachalam2026tomography}}]
A set of states $\mathcal{C}$ is a model class if the following hold:
\begin{enumerate}
    \item Every $n$-qubit state $\ket{\phi} \in \mathcal{C}$ can be described in $\mathsf{poly}(n)$ time.
    \item There is a \textit{weak} agnostic tomography protocol $\mathcal{A}_{\mathsf{WAL}}$ using $\mathcal{S}_{\mathsf{WAL}}$ copies of an unknown state $\ket{\psi}$ and in time $T_{\mathsf{WAL}}$ outputs a $\ket{\phi} \in \mathcal{C}$ such that $|\braket{\phi|\psi}|^2 \geq \eta(F_{\mathcal{C}}(\ket{\psi}))$ where $\eta: [0, 1] \to [0, 1]$.
    \item For each $\ket{\phi}\in \mathcal{C}$, there is a classical procedure $\mathcal{A}_{\mathsf{prep}}$ running in $T_{\mathsf{prep}}$ time that outputs a circuit $V$ that prepares $\ket{\phi}$, i.e., $V\ket{0^n} = \ket{\phi}$ and $V$ has gate complexity $G_{\mathsf{prep}}$.
\end{enumerate}
\label{def:model_class}
\end{definition}
We then define the $\calC$-extent of any quantum quantum state $\ket{\psi}$ as 
\begin{equation}\label{eq:C_extent}
\xi_{\calC}(\ket{\psi}) := \min \{ \norm{c}_1 : \ket{\psi} = \sum_{i} c_i \ket{\phi}_i, \ket{\phi_i} \in \calC \,\, \forall i \}.
\end{equation}
The $\calC$-rank of any quantum state $\ket{\psi}$ is defined over $\norm{c}_0$ in the above definition. The main result of \cite{arunachalam2026tomography} is to provide an algorithm for learning structured decompositions of state over $\calC$.
\begin{theorem}[{\cite[Theorem~1.2]{arunachalam2026tomography}}]\label{thm:model-class-decomposition}
Let $\varepsilon, \delta \in (0,1)$ and $\mathcal{C}$ be a model class as in Definition~\ref{def:model_class}. There is an algorithm that uses copies of an unknown state $\ket{\psi}$ and with probability $\geq 1 - \delta$, outputs a list of $\kappa = \poly(1/\varepsilon,1/\eta(\varepsilon))$ many states $\{\ket{\phi_i}\}_{i \in [\kappa]}$ belonging to $\calC$ and coefficients $\beta \in \mathbb{C}^k$ such that $\ket{\psi}$ can be expressed (up to a global phase) as
$$
\ket{\psi} = \sum_{i = 1}^{\kappa} \beta_i \ket{\phi_i} + \alpha \ket{\phi_R}, \quad \text{where }\alpha^2 \mathcal{F}_{\mathcal{C}} (\ket{\phi_R}) \leq \varepsilon.
$$
The overall complexity of this algorithm is as follows
\begin{align}
& \text{Sample: }\widetilde{O}(\mathcal{S}_{\mathsf{WAL}} \cdot \poly (1/\varepsilon, 1/\eta(\varepsilon), \log (1/\delta))),\\
&\text{Time: }\widetilde{O}(T_{\mathsf{WAL}} \cdot \poly (1/\varepsilon, 1/\eta(\varepsilon), \log (1/\delta)) + S_{\mathsf{WAL}} \cdot \poly (G_{\mathsf{prep}}, T_{\mathsf{prep}}, 1/\varepsilon, 1/\eta(\varepsilon)).
\end{align}
\end{theorem}
Using the above result, they proved the following tomography result for quantum states with bounded extent.
\begin{theorem}[{\cite[Theorem~4.2]{arunachalam2026tomography}}]
\label{thm:bounded-extent-learning}
Let $\varepsilon, \delta \in (0,1)$. Let $\calC$ be a model class as defined in Definition~\ref{def:model_class}. Let $\ket{\psi}$ be an $n$-qubit state with $\xi_{\C} (\ket{\psi}) \leq \xi$. Then, there is an algorithm that with probability $\geq 1 - \delta$ outputs a $\poly(1/(\varepsilon \cdot \eta(\varepsilon)))$ $\calC$-rank state $\ket{\phi}$ such that $\ket{\phi}$ is $\varepsilon$-close to $\ket{\psi}$ in trace distance. The algorithm uses $\widetilde{O}\Big(S_{\mathsf{WAL}} \cdot \poly(\xi, 1/\varepsilon, 1/\eta(\varepsilon), \log(1/\delta))\Big)$ copies of $\ket{\psi}$ and time complexity
$$
\widetilde{O}\Big(T_{\mathsf{WAL}} \cdot \poly(\xi,1/\varepsilon,1/\eta(\varepsilon), \log(1/\delta)) + S_{\mathsf{WAL}} \cdot \poly(G_{\textsf{prep}}, T_{\textsf{prep}}, \xi, 1/\varepsilon, 1/\eta(\varepsilon)) \Big),
$$
where $S_{\textsf{WAL}}, T_{\textsf{WAL}}, T_{\textsf{prep}}, G_{\textsf{prep}}$ are parameters of the model class $\calC$.
\end{theorem}
A corollary of the above theorem is the learnability of states with low stabilizer extent, which we state below.
\begin{corollary}[{\cite[Theorem~1.4]{arunachalam2026tomography}}]\label{corr:bounded-stab-extent-learning}
Let $\xi > 0$, $\varepsilon,\delta \in (0,1)$. Suppose $\ket{\psi}$ is an unknown $n$-qubit state with stabilizer extent $\xi_{\Stab}(\ket{\psi}) \leq \xi$. Then, there exists an algorithm with time complexity $\poly(n (\xi/\varepsilon)^{\log(\xi/\varepsilon)})$ that outputs with probability $\geq 1 -\delta$, a $O(1/\varepsilon^2)$ stabilizer-rank state $\ket{\phi}$ which is $\varepsilon$-close to $\ket{\psi}$ in trace distance.
\end{corollary}

\section{Agnostic tomography of Clifford unitaries}\label{sec:agnostic-learning-cliffords}
In this section, we prove our main result on agnostic tomography of Clifford unitaries given query access to an unknown $n$-qubit unitary $U$. In particular, we will prove Theorem~\ref{thm:agn_learner_cliffords}, which we restate below for convenience.
\agnosticlearnerCliffords*

Our algorithm for the above theorem will crucially use the Choi map and the stabilizer bootstrapping algorithm (Theorem~\ref{thm:stabilizer_bootstrapping}) \cite{chen2025stabilizer} for agnostic tomography of stabilizer states. Consider the set of $2n$-qubit quantums states on subsystems $AB$ where $A$ corresponds to the first $n$ qubits and $B$ corresponds to the last $n$ qubits. Then, Choi states of $n$-qubit Clifford unitaries are exactly the $2n$-qubit stabilizer states with entanglement entropy $n$ across the $AB$ cut. Applying stabilizer bootstrapping directly to $\sket{U}$ (which is guaranteed to have high stabilizer fidelity) may output a $2n$-qubit stabilizer state that does not correspond to the Choi state of an $n$-qubit Clifford i.e., it may not have entanglement entropy $n$ across the $AB$ cut. We thus solve the agnostic tomography of Clifford unitaries problem by solving the more general constrained agnostic stabilizer tomography problem of outputting the nearest $2n$-qubit stabilizer state with entanglement entropy $k$ across the $AB$ cut. To formally state our result, let us define $\calS_k$ as the set of $2n$-qubit stabilizer states with entanglement entropy of $k \in [n]_0$ across $A$ i.e.,
\begin{equation}\label{eq:stab_states_entropyk}
\calS_k := \{ \ket{s} \in \Stab(2n) : E_A(\ket{s}) = k\}.
\end{equation}
We then have the following result.
\begin{theorem}\label{thm:agn_learn_ent_entropy}
Let $\varepsilon,\delta \in (0,1)$ and $k \in [n]_0$. Given copies of an unknown $2n$-qubit quantum state $\ket{\psi}$ with unknown optimal fidelity of $\opt$ with $\calS_k$, there exists an algorithm that outputs $\ket{s} \in \calS_k$ with probability $\geq 1-\delta$ such that
$$
|\la s | \psi \ra|^2 \geq \opt - \varepsilon.
$$
The algorithm uses $\poly(n, (1/\varepsilon)^{\log(1/\varepsilon)}, \log(1/\delta))$ sample and time complexity.
\end{theorem}

\subsection{Stabilizer-neighbor graph}
We now introduce the concept of a stabilizer-neighbor graph which will be important in our main algorithm.

\paragraph{Neighbors of stabilizer states.}
The fidelity between two distinct stabilizer states is at most $1/2$. Motivated by this we define neighborhood between stabilizer states.
\begin{definition}[Stabilizer $k$-neighbors]
Let $\ket{s}$, $\ket{t} \in \Stab(n)$ be stabilizer states. We say that $\ket{s}$ and $\ket{t}$ are \textit{neighbors} if 
\[
    \left|\braket{s|t}\right|^2 = 1/2.
\]
We define the neighborhood of $\ket{s}$ as
\[
N(\ket{s}) = \{ \ket{t} \in \Stab(n)  :  \left|\braket{s|t}\right|^2 = 1/2\}.
\]
Now, let $G$ be the graph whose vertices are stabilizer states, and with edges between two stabilizer states iff they are neighbors.
We say that two stabilizer states $\ket{s}$ and $\ket{t}$ are $k$-neighbors if the shortest path from $\ket{s}$ to $\ket{t}$ in $G$ has length at most $k$.
\label{def:stabilizer-neighbors}
\end{definition}

\begin{fact}\label{fact:neighbor_group_overlap}
Let $\ket{\phi}$ and $\ket{\psi}$ be two neighboring $n$-qubit stabilizer states stabilized by the stabilizer subgroups $S_\phi$ and $S_\psi$ respectively. We have,
$$
|S_\phi \cap S_\psi| = 2^{n-1}.
$$
\end{fact}
\begin{proof}
First, note that any Pauli stabilizing both $\ket\psi$ and $\ket\phi$ must have the same sign in $S_\phi$ and $S_\psi$, since otherwise
\[
    \braket{\psi|\phi} = \bra\psi P (-P \ket\phi) = -\braket{\psi|\phi},
\]
and so
$\left| \braket{\psi|\phi} \right|^2 = 0$, which contradicts that $\ket\psi,\ket\phi$ are neighbors.

Now, note that any stabilizer state $\ket{\psi}$ can be written as a sum of the Pauli operators that stabilize it:
$$
\ketbra{\psi}{\psi} = \frac{1}{2^n} \sum_{P \in S_\psi} P.
$$
We then see that the fidelity between two stabilizer states can then be written as
$$
\left| \braket{\phi|\psi} \right|^2 = \Tr[(\ketbra{\phi}{\phi})(\ketbra{\psi}{\psi})] = \sum_{P \in S_\phi}\sum_{Q \in S_\psi}\frac{1}{2^{2n}}\Tr(PQ) = \frac{1}{2^n} \cdot |S_\phi \cap S_\psi|,
$$
where in the last equality, we used that $\Tr(PQ) = 0$ if $P \neq Q$, otherwise $\Tr(PQ) = 2^n$. Since $\ket{\phi}, \ket{\psi}$ are neighbors, we have that $\left| \braket{\phi|\psi} \right|^2 = \frac{1}{2}$ and thus $\left| S_\phi \cap S_\psi \right| = 2^{n-1}$ as desired. 
\end{proof}

\paragraph{Geometry of $\calS_k$ inside the stabilizer-neighbor graph.}
We now relate the entanglement entropy of neighboring stabilizer states. The following result shows that entanglement entropy changes by at most one on stabilizer-neighbor edges.

\begin{lemma}\label{lem:neighbor_entanglement}
Let $\ket{\phi}, \ket{\psi} \in \Stab(2n)$ be two $2n$-qubit stabilizer states defined across $AB$. If $\ket{\phi},\ket{\psi}$ are neighbors with entanglement entropies on $A$ of $E_A(\ket{\phi})$ and $E_A(\ket\psi)$ respectively, then $\left| E_A(\ket{\phi}) -E_A(\ket{\psi}) \right| \leq 1$.
\end{lemma}
\begin{proof}
Let $S_\phi$, and $S_\psi$ be the stabilizer groups for $\ket{\phi}$, and $\ket{\psi}$ respectively. Let $G := S_\phi \cap S_\psi$. We then have $|G| = |S_\phi \cap S_\psi| = 2^{2n-1}$ by Fact~\ref{fact:neighbor_group_overlap}. Let us choose any element $p \in S_\phi \setminus G$. Let $S_{\phi,A}$ be the set of Pauli operators in $S_\phi$ that act trivially (i.e., action of $\id$) over the subsystem $B$. Similarly, let $G_A$ be the set of Pauli operators in $G$ that act trivially over $B$. We now note that
$$
S_\phi = G \sqcup pG,
$$
where $\sqcup$ denotes the disjoint union. We can write $S_{\phi,A}$ as
\begin{equation}\label{eq:interim1_stab_neighbor}
S_{\phi,A} = (S_{\phi,A}\cap G) \sqcup (S_{\phi,A}\cap pG) = G_A \sqcup (S_{\phi,A} \cap p G) \implies |S_{\phi,A}| = |G_A| + |S_{\phi,A} \cap pG|,
\end{equation}
where we used that $S_{\phi,A}\cap G = G_A$ in the second equality. We now argue that 
\begin{equation}\label{eq:size_SphiA}
|G_A| \leq |S_{\phi,A}| \leq 2|G_A|.
\end{equation}
The first inequality is immediate as $G_A \subseteq S_{\phi,A}$. To obtain the second inequality, we proceed as follows. If $S_{\phi,A}\cap pG = \emptyset$ then we immediately obtain $|S_{\phi,A}|=|G_A|$. If $S_{\phi,A}\cap pG \neq \emptyset$, then let us choose some $h\in S_{\phi,A}\cap pG$. We now claim that $S_{\phi,A}\cap pG = hG_A$. To see this, let us choose another element $s \in S_{\phi,A}\cap pG$ different from $h$. Since both $h$ and $s$ lie in $pG$, their product lies in $G$ i.e., $hs \in G$. Moreover, since both $h$ and $s$ are supported only on $A$, we also have that $hs$ is also only supported on $A$ and hence $hs\in G_A$.
Since $h^2=I$, we then have that $s=h(hs) \in h G_A$. As we can do this for all elements $s \in S_{\phi,A} \cap pG$ that are different from $h$, we then have that
$$
S_{\phi,A}\cap pG \subseteq hG_A.
$$
Conversely, if $g\in G_A$, then $hg$ is supported only on $A$, and since $h\in pG$, we also have $hg\in pG$. Hence,
$$
hG_A\subseteq S_{\phi,A}\cap pG.
$$
Combining the above two observations, we then have that
$$
S_{\phi,A}\cap pG = hG_A \implies |S_{\phi,A}\cap pG| = |G_A|.
$$
This proves Eq.~\eqref{eq:size_SphiA}. Applying the same argument to $S_\psi$ gives
$$
|G_A|\leq |S_{\psi,A}|\leq 2|G_A|.
$$
Combining the inequalities for $|S_{\phi,A}|$ and $|S_{\psi,A}|$, we obtain
$$
\frac{1}{2}
\leq
\frac{|S_{\phi,A}|}{|S_{\psi,A}|}
\leq
2 \implies \left| \log |S_{\phi,A}| - \log |S_{\psi,A}| \right| \leq 1.
$$
Using the stabilizer entropy formula $E_A(|\phi\rangle) = n - \log |S_{\phi,A}|$, we then obtain
$$
|E_A(|\phi\rangle)-E_A(|\psi\rangle)|\leq 1,
$$
which is the desired result. This concludes the proof.
\end{proof}
The above lemma shows that if a path in the stabilizer-neighbor graph starts at a node on a state in $\calS_k$, then after traversing $t$ many edges, the path reaches a node with a stabilizer state of entanglement entropy in $\{k-t,\ldots,k+t\}$.

\subsection{Algorithm and analysis}
We now present our algorithm for agnostic tomography of Clifford unitaries (Algorithm~\ref{algo:strong-agnostic-learner}). The remainder of this section is devoted to its analysis.

\begin{myalgorithm}
\begin{algorithm}[H]\label{algo:strong-agnostic-learner}
\caption{Strong agnostic tomography protocol for stabilizer states with entanglement $k$}
\setlength{\baselineskip}{1.8em} 
\DontPrintSemicolon 
\KwInput{$\varepsilon, \delta \in (0,1)$, sample access to $2n$-qubit state $\ket\psi$ with fidelity $\tau$ to some $2n$-qubit stabilizer state with entanglement entropy $k$ across the $AB$ bipartition.}
\KwOutput{Classical description of $\ket{s} \in \calS_k$ s.t. $\left| \braket{\psi|s} \right|^2 \geq \tau - \varepsilon$.}
\vspace{2mm}
Run the stabilizer bootstrapping algorithm for list-decoding with $\gamma = \frac{1}{2} + \frac{\varepsilon}{12\log (1/\varepsilon)}$ to obtain a list $L$ of stabilizer states of length $O(\log \frac{1}{\delta})$ (\cite{chen2025stabilizer}, Corollary 6.2).  \\
\ForEach{$\ket{s} \in L$}{
    Find local Cliffords $C_A, C_B$ using \cite{fattal2004entanglement} which map $\ket{s}$ into the form
    $\ket{\EPR_{k'}} \otimes \ket{0}^{2(n-k')}$. If $\left| k' - k \right| > 10\log(1/\tau)$, continue to the next $\ket{s}$.

    \If{$k' \geq k$}{
        \ForEach{$s \in \{0,1\}^{k'-k}$}{
             Estimate $E_s = \left| \left( \bra{\EPR_k} \otimes \bra{0}^{2(n-k)} \right) (X^{s} \otimes X^{s})(C_A \otimes C_B) \ket\psi \right|^2$ to additive error $\varepsilon/4$ using Lemma~\ref{lem:shadows_stabilizers}. 
        }
        \textbf{Return} $(C_A^\dagger \otimes C_B^\dagger) (X^{s} \otimes X^{s}) \left( \ket{\EPR_k} \otimes \ket{0}^{2(n-k)} \right)$ corresponding to maximum $E_s$
    }
    \Else{
        \ForEach{$s \in \{0,1\}^{k-k'}$}{
             Estimate $E_s = \left| \left( \bra{\EPR_k} \otimes \bra{0}^{2(n-k)} \right) (\id \otimes Z^{s})(C_A \otimes C_B) \ket\psi \right|^2$ to additive error $\varepsilon/4$ using Lemma~\ref{lem:shadows_stabilizers}. 
        }
        \textbf{Return} $(C_A^\dagger \otimes C_B^\dagger) (\id \otimes Z^{s}) \left( \ket{\EPR_k} \otimes \ket{0}^{2(n-k)} \right)$ corresponding to maximum $E_s$
    }
}
\end{algorithm}
\end{myalgorithm}

The following claim relates the $\mathcal{S}_k$-fidelity to the fidelity of the state outputted by stabilizer bootstrapping. We use a similar idea to the proof of the quasipolynomial time agnostic learner for quadratic phase functions in \cite{briet2026near} (Lemma 4.1).
\begin{claim}\label{claim:bds-t-neighbors}
Let $\tau \in (0,1)$, and let $\gamma \in (\frac12, \frac34]$. Consider the context of Theorem~\ref{thm:agn_learn_ent_entropy}. Suppose the unknown $2n$-qubit state $\ket{\psi}$ has maximal fidelity with $\mathcal{S}_k$ at least $\tau$ and let $\ket{s'} \in \calS_k$ such that $|\la s'|\psi \ra|^2 \geq \tau$. Then with probability at least $((\gamma -1/2 ) \tau)^{O(\log(1/\tau))}$,
stabilizer bootstrapping (Theorem~\ref{thm:stabilizer_bootstrapping}) on $\ket\psi$ will output a state $\ket{s} \in \Stab(2n)$ satisfying the following for some $t \leq 3\log(1/\tau)$:
\begin{itemize}
    \item $\ket{s}$ is a $t$-neighbor of $\ket{s'}$,
    \item $\ket{s}$ has entanglement entropy $k'$, where $|k - k'|\leq t$,
    \item $|\la s | \psi \ra|^2 \geq \tau/\gamma^t$,
    \item $\ket{s}$ is a $\gamma$-approximate local maximizer of fidelity.
\end{itemize}
\end{claim}
\begin{proof}
Either $\ket{s'}$ is a $\gamma$-approximate local maximizer, or $\ket{s'}$ is not a $\gamma$-approximate local maximizer, so there exists $\ket{s_1}$ which is a $1$-neighbor of $\ket{s'}$ such that
\begin{align*}
    \calF(\ket{s'}, \ket{\psi}) < \gamma \calF(\ket{s_1}, \ket{\psi}) \implies \calF(\ket{s_1}, \ket{\psi}) > \dfrac{\tau}{\gamma}.
\end{align*}
Next by the same argument, either $\ket{s_1}$ is a $\gamma$-approximate local maximizer or there is some $\ket{s_2}$ which is a $2$-neighbor of $\ket{s'}$ such that
\begin{align*}
    \calF(\ket{s_1}, \ket{\psi}) < \gamma \calF(\ket{s_2}, \ket{\psi}) \implies \calF(\ket{s_2}, \ket{\psi}) > \dfrac{\tau}{\gamma^2}.
\end{align*}
By \cref{lem:neighbor_entanglement}, we have that each step in the neighbor graph can change entanglement entropy by at most $1$.
Now continuing in this manner, we see there exists a sequence of states $\ket{s_j}$ such that for each $j$,
\begin{itemize}
    \item $\ket{s_j}$ is a $j$-neighbor of $\ket{s'}$.
    \item $\ket{s_j}$ has entanglement entropy within $j$ of $k$.
    \item $\calF(\ket\psi, \ket{s_j}) > \dfrac{\tau}{\gamma^j}$.
\end{itemize}
Finally, since $\dfrac{\tau}{\gamma^j} < \calF(\ket\psi, \ket{s_j}) \leq 1$, the sequence must terminate after $t < \log_{1/\gamma}(1/\tau) \leq 3\log(1/\tau)$ steps. Since the sequence terminates, $\ket{s_t}$ must be a $\gamma$-approximate local maximizer, as desired. Thus, stabilizer bootstrapping is guaranteed to output $\ket{s_t}$ with probability at least $((\gamma -1/2 ) \tau)^{O(\log(1/\tau))}$.
\end{proof}

The crucial observation here is that, although the stabilizer bootstrapping may output a state $\ket{\phi}$ that doesn't have the desired entanglement entropy, we will be compensated with increased fidelity. Thus when searching for a Choi state of a Clifford which is close to $\ket\phi$, we can afford to \textit{lose} some fidelity. Putting these ideas together we get our main result.

\begin{claim}
Let $\tau,\delta, \varepsilon \in (0,1)$, and $0 \leq k \leq n$. Given an unknown $2n$-qubit state $\ket\psi$ that has fidelity $\geq \tau$ with some state in $\mathcal{S}_k$, there is a quantum algorithm that with success probability $\geq 1-\delta$ outputs a state $\ket{s}$ in $\mathcal{S}_k$ such that $|\braket{s|\psi}|^2 \geq \tau-\varepsilon$. The algorithm uses $\poly(n, (\frac{1}{\varepsilon\tau})^{\log(1/\tau)}, \log(1/\delta))$ many copies of $\ket\psi$ and time complexity.
\label{claim:find_entropy_k}
\end{claim}
\begin{proof}

Let $\gamma = 1/2 + \alpha$, where $\alpha \leq \frac{\varepsilon}{12\log(1/\varepsilon)}$. By Claim~\eqref{claim:bds-t-neighbors}, stabilizer bootstrapping on $\ket\psi$ will output a stabilizer state $\ket\phi$ such that for some $t \leq 3\log(1/\tau)$, $\ket\phi$ has entanglement entropy $k'$ such that $|k-k'| \leq t$, and 
\[
\calF(\ket\psi, \ket\phi) \geq \dfrac{\tau}{\gamma^t}.
\]

Since $\ket{\phi}$ is a stabilizer state on the
bipartition $AB$, with $|A|=|B|=n$, we can compute local Clifford unitaries
$C_A,C_B$ (using \cite{fattal2004entanglement}) such that
$$
(C_A \otimes C_B) \ket{\phi} = \ket{\EPR_{k'}} \otimes \ket{0}^{n-k'} \otimes \ket{0}^{n-k'},
$$
where $\ket{\EPR_{k}}$ is a $k$-pair $\EPR$ state across the $AB$ cut. Then
\begin{equation}\label{eq:promise_k}
\frac{\tau}{\gamma^t} \leq \left|\braket{\phi|\psi}\right|^2 = \left|\Big(\bra{\EPR_{k'}} \otimes \bra{0}^{2(n-k')} \Big)(C_A \otimes C_B)\ket{\psi} \right|^2.
\end{equation}

Now, we consider two cases, $k' < k
$, and $k' > k$. First, suppose $k' < k$, and let $a = k-k' \leq t$. Then
\begin{align*}
\ket{\EPR_{k'}} \otimes \ket{0}^{2(n-k')} 
&= 2^{a/2} (\id \otimes \ketbra{0}{0}^{a}) \ket{\EPR_k} \otimes \ket{0}^{2(n-k)} \\
&= 2^{a/2} (\id \otimes \dfrac{1}{2^a}\sum_{s \in \{0,1\}^a} Z^s) \ket{\EPR_k} \otimes \ket{0}^{2(n-k)} \\
&= 2^{a/2} \mathbb{E}_{s} \left[ \left( \id \otimes Z^s \right) \ket{\EPR_k} \otimes \ket{0}^{2(n-k)} \right].
\end{align*}
Thus, there exists $s^*$ such that 
\begin{align*}
\left| \left( \bra{\EPR_k} \otimes \bra{0}^{2(n-k)} \right) (\id \otimes Z^{s^*})(C_A \otimes C_B) \ket\psi \right|^2 \geq \dfrac{\tau}{\gamma^t 2^a} \geq \dfrac{\tau}{2^t\gamma^t}.
\end{align*}

Next, suppose $k' > k$, and let $a = k'-k \leq t$. Then
\begin{align*}
\ket{\EPR_{k'}} \otimes \ket{0}^{2(n-k')} 
&= \ket{\EPR_k} \otimes \ket{\EPR_a} \otimes \ket{0}^{2(n-k')} \\
&= \ket{\EPR_k} \otimes \left( \dfrac{1}{2^{a/2}}\sum_{s \in \{0,1\}^a} \ket{s}\ket{s} \right) \otimes \ket{0}^{2(n-k')} \\
&= 2^{a/2} \mathbb{E}_s \left[ \ket{\EPR_k} \otimes (X^s \otimes X^s) \ket{0}^{2a} \otimes \ket{0}^{2(n-k')} \right].
\end{align*}

Thus by a probabilistic argument, there exists $s^*$ such that 
\begin{align*}
\left| \left( \bra{\EPR_k} \otimes \bra{0}^{2(n-k)} \right) (X^{s^*} \otimes X^{s^*})(C_A \otimes C_B) \ket\psi \right|^2 \geq \dfrac{\tau}{\gamma^t 2^a} \geq \dfrac{\tau}{2^t\gamma^t}.
\end{align*}

In both cases, since applying local Cliffords does not change the entanglement entropy, we can find by brute force over $2^{t} \leq \frac{1}{\tau^{3}}$ bitstrings a stabilizer state $\ket\phi \in \mathcal{S}_k$ such that 
\[
\left| \braket{\psi|\phi} \right|^2 \geq \dfrac{\tau}{2^t \gamma^t} = \dfrac{\tau}{(1+2\alpha)^t}.
\]

Now since $2\alpha t \leq \frac{\varepsilon}{2}$, we have $(1+2\alpha)^t \leq e^{2\alpha t} \leq e^{\varepsilon/2}$. Thus
$$
\left| \braket{\phi|\psi} \right|^2 \geq \dfrac{\tau}{(1+2\alpha)^t} \geq e^{-\varepsilon/2}\tau \geq \left(1 - \frac{\varepsilon}{2}\right)\tau \geq \tau - \frac{\varepsilon}{2}.
$$

Thus, by estimating $\left| \braket{\phi|\psi} \right|^2$ to additive error of $\frac{\varepsilon}{4}$, we can identity a stabilizer state with fidelity at least $\tau - \varepsilon$.

The complexity of the algorithm is due to using \cite{chen2025stabilizer} which requires $\poly(n,(1/\varepsilon\tau)^{\log(1/\tau)}, \log(1/\delta))$
time complexity and determining $s^\star$ which requires $\poly(1/\tau, \log(1/\delta))$ samples and $\poly(n,1/\tau, \log(1/\delta))$ time complexity.
\end{proof}
\noindent A special case of \cref{claim:find_entropy_k} allows us to learn the nearest Clifford operator to a given unitary.

\begin{corollary}\label{cor:clifford_tau}
Let $\tau,\delta,\varepsilon \in (0,1)$. Given an unknown $n$-qubit unitary $U$ that has Clifford fidelity $\geq \tau$, there is a quantum algorithm that with success probability $\geq 1-\delta$ outputs a Clifford unitary $V$ such that $|\la\!\la U | V \ra\!\ra|^2 \geq \tau-\varepsilon$. The algorithm uses $\poly(n, (\frac{1}{\varepsilon\tau})^{\log(1/\tau)}, \log(1/\delta))$ many queries to $U$ and time complexity.
\end{corollary}
\begin{proof}
    Let $\mathcal{A}$ be the algorithm from \cref{claim:find_entropy_k}. We simply set $k = n$ and apply $\mathcal{A}$ on the unitary $U$ to output a stabilizer state with entanglement entropy $n$ of the form 

    \begin{equation}
        (C_A \otimes C_B) P \ket{\EPR_k} = (C_{A'} \otimes C_{B'}) \ket{\EPR_k} = (C_{A'}C_{B'}^{\top} \otimes \id)\ket{\EPR},
    \end{equation}
    where $P$ is some $2n$-qubit Pauli operator and $C_A, C_B, C_{A'}, \text{and } C_{B'}$ are $n$-qubit Clifford operators. Outputting $V = C_{A'}C_{B'}^{\top}$ yields the desired Clifford unitary.
\end{proof}

The proof of Theorem~\ref{thm:agn_learner_cliffords} then follows from Corollary~\ref{cor:clifford_tau} by binary search over the choice of $\tau$ to handle the unknown optimal fidelity $\opt$. In particular, let $\tau$ be a guess for $\opt$, which is used as an input to Algorithm~\ref{algo:strong-agnostic-learner}. If $\tau > \opt$, the algorithm will fail to find a Clifford with the desired fidelity. Thus, one can decrease $\tau$ until the algorithm returns a Clifford with fidelity at least $\tau - \varepsilon$. At this point, $\opt$ is known within a multiplicative factor of $2$, and a grid search over this range using increments of size $O(\varepsilon)$ will determine $\opt$ to the desired precision.

\section{Applications}
\label{sec:applications}
In this section we give some applications of our strong agnostic tomography protocol.

\subsection{Learning unitaries with bounded Clifford extent}
We now describe how agnostic tomography of Clifford unitaries (Theorem~\ref{thm:agn_learner_cliffords}) may be used to perform tomography of unitaries with bounded Clifford extent. We first show that for arbitrary unitaries, we can learn a structured Clifford decomposition of $U$.

\paragraph{On learning Clifford decompositions.}
We will invoke Theorem~\ref{thm:model-class-decomposition} to show this result. Let $\C$ denote the set of Choi states of $n$-qubit Clifford unitaries. We note that $\C \subset \Stab(2n)$ and any $2n$-qubit stabilizer state can be succinctly described by $2n$ generators of its stabilizer subgroup. Furthermore there exists an (efficient) classical procedure for outputting a circuit preparing such a state as stabilizer states are known to efficiently prepareable. 
\begin{lemma}[Clifford synthesis~\cite{dehaene2003clifford,patel2003efficient}]\label{lem:clifford_synthesis}
Given the classical description of an $n$-qubit stabilizer state $\ket{\phi}$, there is a classical algorithm that outputs a Clifford circuit $\textsf{C}$ that prepares $\ket{\phi}$, using $O(n^2)$ many single-qubit and two-qubit Clifford gates.
\end{lemma}
Taken together, with our agnostic tomography protocol for Choi states of Cliffords (Theorem~\ref{thm:agn_learn_ent_entropy} with $k$ there set as $n$)\footnote{We use this theorem as the input unknown state $\ket{\psi}$ can be a $2n$-qubit state and we require our agnostic learner to accept such input states in Theorem~\ref{thm:model-class-decomposition}.}, shows that $\C$ is a model class. We then obtain the following from \Cref{thm:model-class-decomposition}.
\begin{corollary}
\label{cor:clifford_decomposition}
Let $\varepsilon,\delta \in (0,1)$. Given query access to an unknown $n$-qubit unitary $U$, there is a quantum algorithm that with success probability $\geq 1 - \delta$ outputs a list of $\kappa = O(1/\varepsilon^2)$ many Clifford unitaries $\{V_i\}_{i \in [\kappa]}$ and corresponding coefficients $\{\beta_i\}_{i \in [\kappa]}$ such that $\sket{U}$ can be expressed (up to a global phase) as
$$
\sket{U} = \sum_{i=1}^\kappa \beta_i \sket{V_i} + \alpha \ket{\psi},
$$
where $\ket{\psi}$ is a $2n$-qubit state satisfying $\alpha^2 \cdot \max_{V \in \Cliff(n)} |\la \psi \sket{V}|^2 \leq \varepsilon$. The algorithm uses query complexity and time complexity $\poly(n, (1/\varepsilon)^{\log(1/\varepsilon)}, \log (1/\delta))$.    
\end{corollary}
We remark that although the above corollary is given for Choi states of arbitrary $n$-qubit unitaries, it can be shown for any $2n$-qubit state.

\paragraph{Tomography result.}
When $U$ is promised to have Clifford extent $\leq \xi$, we note that $\sket{U}$ will have extent $\leq \xi$ with respect to the class $\C$ of Choi states of Cliffords. We then obtain the following tomography protocol for unitaries with bounded Clifford extent from Theorem~\ref{thm:bounded-extent-learning}.
\begin{corollary} \label{cor:learning_low_clifford_extent}
Let $\varepsilon,\delta \in (0,1)$ and $\xi \geq 1$. Suppose $U$ is an unknown $n$-qubit unitary with Clifford extent $\leq \xi$. Then, there exist an algorithm that given query access to $U$ and with probability $\geq 1-\delta$, outputs an operator $V$ with Clifford rank $O(\xi^2/\varepsilon^2)$ such that
$$
\dist(U,V) \leq \varepsilon,
$$
where $\dist$ is as defined in Definition~\ref{def:dist_linear_ops}. The query and time complexity of this algorithm is $\poly(n, (\xi/\varepsilon)^{\log(\xi/\varepsilon)}, \log (1/\delta))$.
\end{corollary}
We remark that the above protocol is not proper as it outputs an operator that is not necessarily even unitary. We leave the task of obtaining a proper tomography protocol as future work.

\subsection{Learning Hamiltonians with Clifford structure}
\label{sec:Hamiltonian-learning}
\subsubsection{Hamiltonians with bounded Clifford extent}
We now consider the problem of tomography for Hamiltonians with bounded Clifford extent or rank (\cref{def:extent/rank}). In particular, we have the following main result.
\begin{theorem}\label{thm:HL_low_clifford_extent}
Let $\varepsilon,\delta \in (0,1)$ and $\xi \geq 1$. Suppose $H$ is an unknown $n$-qubit Hamiltonian with Clifford extent at most $\xi$ and satisfies $\Tr(H)=0$. Then, there is an algorithm that given query access to $\exp(-iHt)$ outputs $\widehat{H}$ that has Clifford rank $O(\xi^4/\varepsilon^4)$ with probability $\geq 1-\delta$ such that
$$
\norm{\widehat{H} - H}_{\overline 2} \leq \varepsilon.
$$
The algorithm uses $t=O(\varepsilon/\xi^2)$ and total time evolution of $\poly(n (\xi/\varepsilon)^{\log(\xi/\varepsilon)}\log(1/\delta))$. The algorithm uses query complexity and time complexity $\poly(n (\xi/\varepsilon)^{\log(\xi/\varepsilon)}\log(1/\delta))$.
\end{theorem}
Note that we learn under the normalized Forbenius norm which is given by $\|H\|_{\overline 2} := \sqrt{\frac{\text{Tr}[H^\dagger H]}{2^n}}$. To obtain the above result, we will use Algorithm~\ref{algo:HL_clifford_extent} below that in turn leverages the tomography algorithm (\cref{cor:learning_low_clifford_extent}) for unitaries with bounded Clifford extent.  

\begin{myalgorithm}
\begin{algorithm}[H]\label{algo:HL_clifford_extent}
\caption{Tomography algorithm for Hamiltonians with low Clifford extent}
\setlength{\baselineskip}{1.7em} 
\DontPrintSemicolon 
\KwInput{$\varepsilon, \delta \in (0,1)$, query access to $\exp(-iHt)$ where $H$ has Clifford extent $\leq \xi$.}
\KwOutput{$\widehat{H} = \sum_{i=1}^K \beta_i C_i$ described by list of Clifford unitaries $\{C_i\}_{i \in [K]}$ and corresponding coefficients $\{\beta_i\}_{i \in [K]}$}
\vspace{2mm}
Set time-evolution $t = \varepsilon/(3\xi^2 e)$ and $U(t) = \exp(-iHt)$. \\
Set error parameter $\varepsilon_1 = \varepsilon t/(6 \sqrt{2}). $ \\
Apply $U(t) \otimes \id_n$ to $\ket{\EPR_n}$ to obtain the Choi state $\sket{\exp(-iHt)}.$ \\
Learn an estimate $V$ of $U(t)$ such that $\dist(V,U(t)) \leq \varepsilon_1$ using Corollary~\ref{cor:learning_low_clifford_extent}. \\
Set $\phi = \arg \Tr(V)$ and then $\widehat{U}(t) = e^{- i\phi} V$. \\
Set $\widetilde{H} = (\id - \widehat{U}(t))/(it)$ and then $\widehat{H} = (\widetilde{H} + \widetilde{H}^\dagger)/2$. \\
\textbf{Return} $\widehat{H}$
\end{algorithm}
\end{myalgorithm}
We are now ready to give the proof of Theorem~\ref{thm:HL_low_clifford_extent}.
\begin{proof}[Proof of Theorem~\ref{thm:HL_low_clifford_extent}]
We will use Algorithm~\ref{algo:HL_clifford_extent} to learn $H$. Let the Clifford decomposition of the unknown Hamiltonian $H$ be given by 
\begin{equation}\label{eq:cliff_decomp_H}
    H = \sum_{i=1}^M \alpha_i C_i,
\end{equation}
where $C_i$ are Clifford unitaries for all $i \in [M]$ for some $M \in \mathbb{N}$. We first note that the norm of $H$ can be upper bounded in terms of the extent $\xi$ as
\begin{align}\label{eq:normH_2}
\norm{H}_{\overline 2} = \norm{\sum_i \alpha_i C_i }_{\overline 2}\leq \sum_i |\alpha_i| \norm{C_i}_{\overline 2} \leq \sum_i |\alpha_i| \leq \xi,
\end{align}
where we used the Clifford decomposition of $H$ in the first equality, the traingle inequality in the second inequality, the fact that $\norm{C_i}_2^2 = \Tr[C_i^\dagger C_i]/2^n = \Tr[\id]/2^n = 1$ in the third inequality, and used the definition of extent in the last line. Likewise, since unitaries have operator norm $1$ and thus $\norm{C_i}_{\mathrm{op}} = 1, \forall i \in [M]$, we can show that
\begin{align}\label{eq:normH_op}
\norm{H}_{\mathrm{op}} \leq \xi.
\end{align}

Let $t \in (0,1/2)$ to be fixed later. Let us denote $U(t) = \exp(-iHt)$. Using the Taylor expansion of $U(t)$, we can write
\begin{equation}\label{eq:taylor_expansion_U}
U(t) = \id - itH + \underbrace{R_t}_{\text{higher order terms}}
\end{equation}
where the Taylor remainder $R_t = \sum_{k=2}^\infty (-itH)^k/k!$. To learn $H$, our approach is to learn an approximation $\widehat{U}(t)$ of $U(t)$ and then set the approximation of $H$ to be $(\id - \widehat{U}(t))/(it)$. To learn $\widehat{U}(t)$ of the unitary $U(t)$, we first argue that $U(t)$ itself has bounded Clifford extent. Since $H$ has the Clifford decomposition of Eq.~\eqref{eq:cliff_decomp_H}, we note that for $k \in \mathbb{N}$
$$
H^k = \sum_{i_1,\ldots,i_k \in [M]} \alpha_{i_1} \ldots \alpha_{i_k} C_{i_1} \ldots C_{i_k}.
$$
As the product of Clifford unitaries is another Clifford unitary, we can then express $H^k$ as
$$
H^k = \sum_{i_1,\ldots,i_k \in [M]} \alpha_{i_1} \ldots \alpha_{i_k} \widetilde{C}_{i_1, \ldots i_k},
$$
where we have defined $\widetilde{C}_{i_1, \ldots i_k} := C_{i_1} \ldots C_{i_k}$. Hence, the Clifford decomposition of $U(t)$ is 
\begin{equation}
U(t) = \id + \sum_{k=1}^\infty \frac{(-it)^k}{k!}H^k = \id + \sum_{k=1}^\infty \frac{(-it)^k}{k!} \left[ \sum_{i_1,\ldots,i_k \in [M]} \alpha_{i_1} \ldots \alpha_{i_k} \widetilde{C}_{i_1, \ldots i_k} \right].
\end{equation}
Its corresponding Clifford extent, which we denote as $\xi_{\Cliff}(U(t))$, can then be upper bounded as 
\begin{equation}\label{eq:bound_cliff_extent_sketU}
\xi_{\Cliff}(U(t)) \leq \sum_{k=0}^\infty \frac{t^k}{k!} \left( \sum_{j=1}^M |\alpha_j| \right)^k \leq \sum_{k=0}^\infty \frac{t^k \xi^k}{k!} = e^{t \xi}.
\end{equation}
We will choose $t$ such that $t \leq 1/\xi$. This ensures that $\xi_{\Cliff}(U(t)) \leq e$. Let $\varepsilon_1 \in (0,1)$ be an error parameter to be fixed later. At this point, we can then apply Corollary~\ref{cor:learning_low_clifford_extent} to $U(t)$ with Clifford extent there instantiated as $e$ and error parameter $\varepsilon_1$, which outputs an operator $V$ such that
\begin{equation}\label{eq:promise_V_in_HL}
\norm{V - e^{i\theta} U(t)}_{\overline 2} \leq \sqrt{2} \varepsilon_1,
\end{equation}
for some $\theta \in [0,2\pi)$ and where $V$ is given to us as linear combination of $\kappa = O(1/\varepsilon_1^2)$ many Clifford unitaries $\{C_j\}_{j \in [\kappa]}$ with coefficients $\{\beta_j\}_{j \in [\kappa]}$ i.e., $V =  \sum_{j=1}^\kappa \beta_j C_j$. 

To output an estimate $\widehat{H}$ of $H$, we first define $\phi := \arg \Tr(V)$, set $\widehat{U}(t):=e^{-i\phi} V$ and then set
\begin{equation}\label{eq:tildeH}
\widetilde{H} := (\id - \widehat{U}(t))/(it),
\end{equation}
and since $\widetilde{H}$ may not be Hermitian, we set
\begin{equation}\label{eq:hatH}
    \widehat{H} = (\widetilde{H} + \widetilde{H}^\dagger)/2.
\end{equation}
We will now argue that $\widehat{H}$ is close to $H$ by directly evaluating
\begin{align}
\norm{\widehat{H} - H}_{\overline 2} = \norm{\frac{\widetilde{H} + \widetilde{H}^\dagger}{2} - \frac{H + H^\dagger}{2}}_{\overline 2} \leq \norm{\widetilde{H} - H}_{\overline 2} 
&= \norm{ \frac{\id - \widehat{U}(t)}{it} - \frac{\id - U(t) + R_t}{it}}_{\overline 2} \\
&\leq \frac{\| \widehat{U}(t) - U(t)\|_{\overline 2} + \|R_t\|_{\overline 2}}{t}, \label{eq:promise_hatU}
\end{align}
where we used $H = H^\dagger$ in the first equality, $\norm{\widetilde{H}^\dagger - H^\dagger} = \norm{\widetilde{H} - H}$ along with the triangle inequality in the second inequality, the definition of $\widetilde{H}$ (Eq.~\eqref{eq:tildeH}) and the Taylor expansion of $U(t)$ (Eq.~\eqref{eq:taylor_expansion_U}) in the third inequality, and the triangle inequality in the second line. We first upper bound $R_t$ as 
\begin{align}\label{eq:ub_Rt}
\|R_t\|_{\overline 2} = \left\|\sum_{k=2}^{\infty}\frac{(-itH)^k}{k!}\right\|_{\overline 2} \leq \sum_{k=2}^{\infty} \frac{t^k\|H^k\|_{\overline 2}}{k!} \leq  \sum_{k=2}^{\infty} \frac{t^k \xi^k}{k!} = e^{t \xi} - 1 - t \xi \leq \frac{t^2 \xi^2}{2} e^{t \xi} \leq \frac{t^2 \xi^2}{2} e,
\end{align}
where we used $\norm{H^k}_{\overline 2} \leq \norm{H}_{\mathrm{op}}^{k-1} \norm{H}_{\overline 2} \leq \xi^k$ from Eqs~\eqref{eq:normH_2}--\eqref{eq:normH_op} in the third inequality, $e^x - 1 - x \leq (x^2/2)e^x, \forall x \geq 0$ in the fifth inequality and used the fact that $t\leq 1/\xi$ in the last inequality. 

We now upper bound $\| \widehat{U}(t) - U(t)\|_{\overline 2}$. We first note that 
\begin{align}
\| \widehat{U}(t) - U(t)\|_{\overline 2} = \| e^{-i \phi} V - U(t)\|_{\overline 2} &=   \| e^{-i \phi} (V - e^{i \theta} U(t)) + (e^{i(\theta - \phi)} - 1)U(t) \|_{\overline 2} \nonumber \\
&\leq   \| V - e^{i \theta} U(t) \| + |e^{i(\theta - \phi)} - 1| \cdot \|U(t) \|_{\overline 2} \nonumber \\
&\leq  \sqrt{2}\varepsilon_1 + |e^{i(\theta - \phi)} - 1|, \label{eq:bound_hatU_U}
\end{align}
where we used Eq.~\eqref{eq:promise_V_in_HL} in the third line along with the fact that $\|U(t) \|_{\overline 2}=1$ for a unitary $U(t)$. To bound the second term in Eq.~\eqref{eq:bound_hatU_U}, we first note that
using Fact~\ref{fact:normalized_trace} along with Eq.~\eqref{eq:promise_V_in_HL} gives us
\begin{equation}\label{eq:trace_bound}
\Big| \frac{\Tr(V)}{2^n} - e^{i \theta} \frac{\Tr(U(t))}{2^n}\Big| \leq \sqrt{2} \varepsilon_1.
\end{equation}
From Eq.~\eqref{eq:taylor_expansion_U} and using $\Tr(H)=0$, we have that
\begin{equation}\label{eq:trace_bound2}
\frac{\Tr(U(t))}{2^n} = 1 + \frac{\Tr(R_t)}{2^n} \implies \Big|\frac{\Tr(U(t))}{2^n} - 1\Big| \leq \norm{R_t}_{\overline 2}.    
\end{equation}
Let us denote $z:= \Tr(V)/2^n$. We have
\begin{equation}
|z - e^{i\theta}| = \Big|z - e^{i\theta}\frac{\Tr(U(t))}{2^n} + e^{i \theta} \left[\frac{\Tr(U(t))}{2^n} - 1\right] \Big| \leq \Big|z - e^{i\theta}\frac{\Tr(U(t))}{2^n}\Big| + \Big| \frac{\Tr(U(t))}{2^n} - 1 \Big| \leq \sqrt{2}\varepsilon_1 + \norm{R_t}_{\overline 2},
\end{equation}
where we used $|e^{i \theta}|=1$ in the second inequality followed by Eq.~\eqref{eq:trace_bound} and Eq.~\eqref{eq:trace_bound2} in the final inequality. Let $\Delta := \sqrt{2}\varepsilon_1 + \norm{R_t}_{\overline 2}$. We now note that
\begin{align}
|e^{i(\theta - \phi)} - 1| = |e^{i\phi} - e^{i\theta}| \leq |e^{i \phi} - z| + |z - e^{i \theta}| 
&= \Big|\frac{z}{|z|} - z\Big| +  \Delta \\
&\leq \Big||z| - 1\Big| + \Delta \\
&\leq |z - e^{i \theta}| + \Delta \\
&\leq 2 \Delta, \label{eq:bound_term2}
\end{align}
where we used the definition of $\phi = \arg \Tr(V) = \arg z$ in the third inequality in the first line along with Eq.~\eqref{eq:trace_bound2}, used $| |a| - |b| | \leq |a-b|$ for complex numbers $a,b$ (with $a=z$ and $b=e^{i\theta}$) in the third line, and finally Eq.~\eqref{eq:trace_bound2} in the last line. Substituting Eq.~\eqref{eq:bound_term2} into Eq.~\eqref{eq:bound_hatU_U} and the bound on $\norm{R_t}_{\overline 2}$ from Eq.~\eqref{eq:ub_Rt} gives us
$$
\| \widehat{U}(t) - U(t)\|_{\overline 2} \leq 3 \sqrt{2} \varepsilon_1 + t^2 \varepsilon^2 e.
$$
Substituting the above equation and Eq.~\eqref{eq:ub_Rt} back into Eq.~\eqref{eq:promise_hatU} gives us
$$
\norm{\widehat{H} - H}_{\overline 2} \leq \frac{3 \sqrt{2} \varepsilon_1}{t} + \frac{3 t \xi^2 e}{2}.
$$
Setting $t = \varepsilon/(3 e \xi^2)$ and $\varepsilon_1 = \varepsilon t/(6 \sqrt{2}) = \varepsilon^2/(18\sqrt{2} e \xi^2)$ gives us the desired result of 
$$
\norm{\widehat{H} - H}_{\overline 2} \leq \varepsilon.
$$
The query complexity is due to application of Corollary~\ref{cor:learning_low_clifford_extent} to learn $V$ which is $\poly(n (1/\varepsilon_1)^{\log(1/\varepsilon_1)}) = \poly(n (\xi/\varepsilon)^{\log(\xi/\varepsilon)})$. The total time evolution is then $\poly(n (\xi/\varepsilon)^{\log(\xi/\varepsilon)})$. The time complexity associated with using Corollary~\ref{cor:learning_low_clifford_extent} is $\poly(n (\xi/\varepsilon)^{\log(\xi/\varepsilon)})$. The classical time complexity in computing $\arg \Tr(V)$ is $O(\kappa n^3) = \poly(n \xi/\varepsilon)$ since $\Tr(V) = \sum_j \beta_j \Tr(C_j)$ and $\Tr(C_j)/2^n = \la\!\la \id | C_j \ra\!\ra$ which can be computed in $O(n^3)$ time using the classical algorithm to compute stabilizer inner products from \cite{garcia2017geometry}. The overall time complexity of the algorithm is $\poly(n (\xi/\varepsilon)^{\log(\xi/\varepsilon)})$. This completes the proof.
\end{proof}

\paragraph{Learning Hamiltonians with bounded Clifford rank.} The above protocol can also be utilized to learn Hamiltonians that are promised to have bounded Clifford rank. This is summarized in the result below.
\begin{corollary}
Let $\varepsilon,\delta \in (0,1)$, $\beta > 0$ and $\kappa \in \mathbb{N}$. Suppose $H$ is an unknown $n$-qubit Hamiltonian with Clifford rank at most $\kappa$, has bounded norm of $\norm{H}_{\mathrm{op}} \leq \beta$ and satisfies $\Tr(H)=0$. Then, there is an algorithm that given query access to $\exp(-iHt)$ outputs $\widehat{H}$ that has Clifford rank $O(\beta^4 \kappa^{\kappa^4}/\varepsilon^4)$ with probability $\geq 1-\delta$ such that
$$
\norm{\widehat{H} - H}_{\overline 2} \leq \varepsilon.
$$
The algorithm uses $t=O\Big(\varepsilon/(\beta \kappa^\kappa)^2\Big)$ and total time evolution of $\poly(n (\beta \kappa^{\kappa}/\varepsilon)^{\kappa \log(\beta \kappa/\varepsilon)}\log(1/\delta))$. The algorithm uses query complexity and time complexity $\poly(n (\beta \kappa^{\kappa}/\varepsilon)^{\kappa \log(\beta \kappa/\varepsilon)}\log(1/\delta))$.
\end{corollary}
\begin{proof}
We will show that $H$ having bounded Clifford rank and norm implies that $\sket{\exp(-iHt)}$ has bounded stabilizer extent for a suitable choice of $t \in (0,1)$. We can then use the tomography protocol described in the proof of Theorem~\ref{thm:HL_low_clifford_extent} and the result then follows.

Let the minimal Clifford rank decomposition of $H$ be
$$
H = \sum_{i=1}^\kappa \alpha_i C_i, \quad \text{ where } C_i \in \Cliff(n), \,\, \forall i \in [k].
$$
Let $\sket{H} := (H \otimes \id) \ket{\EPR_n}$. We note that $\norm{\sket{H}}_2 = \norm{H}_2/\sqrt{2^n} \leq \norm{H}_{\mathrm{op}} \leq \beta$. Let us then define
\begin{equation}
    \sket{\widetilde{H}} := \frac{\sket{H}}{\beta} = \sum_{i=1}^\kappa \frac{a_i}{\beta} \sket{C_i} 
\end{equation}
which is a valid pure normalized quantum state and has the stabilizer decomposition as given in the final equality after noting that each $\sket{C_i}$ is a stabilizer state. It then follows from Theorem~\ref{thm:ub_stab_extent_stab_rank_states} that
\begin{equation}\label{eq:ub_stab_rank_tildeH}
\xi_{\Stab}(\sket{\widetilde{H}}) \leq \sum_{i=1}^k |\alpha_i|/\beta \leq \kappa^{O(\kappa)} \implies \sum_{i=1}^\kappa |\alpha_i| \leq \beta \kappa^{O(\kappa)}.
\end{equation}
This shows that the Clifford extent of $H$ is bounded as $\xi_{\Cliff}(H) \leq \beta \kappa^{O(\kappa)}$. Using Theorem~\ref{thm:HL_low_clifford_extent} and the corresponding algorithm with this bound on the Clifford extent gives us the desired result. This completes the proof.
\end{proof}

\subsubsection{Bounds based on stabilizer decomposition of $\sket{H}$}
The tomography protocol above learns a Hamiltonian that is promised to have bounded Clifford extent. However, it is possible that a quantum Hamiltonian does not have small Clifford extent but instead its corresponding normalized Choi state $\sket{H}$ has a small stabilizer extent. We now focus of the family of Hamiltonians that can be represented as
\begin{equation}
H = \sum_{j,k = 1}^R h_{j,k} \ket{s_j} \bra{s_k}
\label{eq:Hamiltonian-family},    
\end{equation}
where $\ket{s_j}$ are stabilizer states with Gram matrix ${\bf G}$ with entries $G_{jk} = \braket{s_j|s_k}$ and ${\bf h} \in \mathbb{C}^{R \times R}$ is a Hermitian matrix. 

We first study the performance of \Cref{thm:HL_low_clifford_extent} on this instance. We note that each term $\ket{s_j}\bra{s_k} = C_j \ket{0^n}\bra{0^n} C_k^\dagger$ for some Clifford unitaries $C_j$ and $C_k$. Therefore 
$$
\xi_{\Cliff} (H) \leq (\sum_{j,k} |h_{j,k}|) \xi_{\Cliff} (\ket{0^n}\bra{0^n})
$$
since
$$
\ketbra{0^n}{0^n} = \frac{1}{2^n} \sum_{S \subseteq [n]} Z^S
$$
we find that
$\xi_{\Cliff} (\ket{0^n}\bra{0^n}) \leq 1$ and therefore
$$
\xi_{\Cliff} (H) \leq (\sum_{j,k} |h_{j,k}|) \leq R \sqrt{\Tr (\mathbf{h}^2)} \leq R^{3/2} \|\mathbf{h}\|_{\rm op}
$$
where in the second inequality we have used Cauchy-Schwartz and in the third inequality we used the relationship between Schatten $2$ and $\infty$ norms. As a result, sample and time complexity of the algorithm in \Cref{thm:HL_low_clifford_extent} to learn $H$ within $\varepsilon$ normalized Frobenius norm with probability at least $\geq 1 - \delta$ scales as $\poly(n (R/\varepsilon)^{\log(R/\varepsilon)}\log(1/\delta))$. We can indeed find instances for which $\|{\bf h}\|_{\rm op} = O(1)$ but $\sum_{j,k} |h_{j,k}|$ scales with $R$. The simplest example is ${\bf h} = I_R$. We can indeed do something stronger and show that there are instances of ${\bf h}$ for which $\|{\bf h}\|_{\rm op} = 1$ but $\sum_{j,k} |h_{j,k}| = R^{3/2}$ (saturating the upper bound above). To see this take ${\bf S}$ to be a symmetric Hadamard matrix in $\{\pm 1\}^{R \times R}$ such that ${\bf S}^T {\bf S}= R I_R$ and let ${\bf h} = \frac{\bf S}{\sqrt{R}}$. ${\bf h}^2 = I_R$ therefore $\|{\bf h}\|_{\rm op} = 1$. Nevertheless, $\sum_{j,k} |h_{j,k}| = R^{3/2}$. 

We now give an alternative algorithm based on small stabilizer extent for $\sket{H}$ directly using time and sample complexity that does not directly scale with $R$ so long as $R \ll 2^{n/3}$.

\begin{theorem}[Hamiltonian tomography based on stabilizer decomposition] Let $H$ be an $n$-qubit Hamiltonian from the family specified in \Cref{eq:Hamiltonian-family} such that $\Tr(H)=0$, $R \leq 2^{n/3}$ and with parameters
$$
\|\mathbf{h}\|_{\rm op} = h^*, \quad \|{\bf G}\|_{\rm op} = g^*.
$$
Then, there is an algorithm using query access to $\exp(-itH)$ which outputs with probability at least $1-\delta$, an efficient representation $\widehat{H}$ such that
$$
\norm{H - \widehat H}_{\overline 2} \leq \varepsilon. 
$$
The algorithm uses $\poly\Big(n (g^*h^*/\varepsilon)^{\log (g^*h^*/\varepsilon)} \log(1/\delta)\Big)$ query and time complexity. The algorithm uses $t=O(\varepsilon/(g^* h^*)^2)$ and the total time evolution is $\poly\Big(n (g^*h^*/\varepsilon)^{\log (g^*h^*/\varepsilon)} \log(1/\delta)\Big)$.
\end{theorem}
\begin{proof}
Let
$$
{\bf h}^{(k)} := {\bf h} ({\bf Gh})^{k-1}.
$$
Then 
$$
H^\ell = \sum_{j,k = 1}^R h^{(\ell)}_{j,k} \ket{s_j}\bra{s_k}
$$
similar to the Hamiltonian tomography procedure in \Cref{thm:HL_low_clifford_extent} we first bound the stabilizer extent of $\sket{e^{- i Ht}}$ using its Taylor expansion as follows
$$
\sket{e^{- i Ht}} = \ket{\EPR_n} + \frac{1}{\sqrt{2^n}} \sum_{\ell = 1}^\infty \frac{(-it)^\ell}{\ell !} \left[\sum_{j,k = 1}^R h^{(\ell)}_{j,k}\ket{s_j}\ket{s^*_k} \right],
$$
which implies
\begin{equation}\label{eq:interim1_stab_extent_Ut}
\xi_{\Stab} (\sket{e^{- i Ht}}) \leq  1+ \frac{1}{\sqrt{2^n}} \sum_{\ell = 1}^\infty \frac{|t|^\ell}{\ell !} \left[\sum_{j,k = 1}^R   |h^{(\ell)}_{j,k}|\right],    
\end{equation}
where we have noted that $\ket{\EPR_n}$ and $\ket{s_j}\ket{s_k^*} \forall j,k$ are stabilizer states. We can then show 
\begin{claim}\label{claim:stab-extent-bound-Ut}
Let $R \leq 2^{n/3}$, $g^* := \|{\bf G}\|_{\rm op}$ and $h^* := \|{\bf h}\|_{\rm op}$, then for $t \geq 0$
$$
\xi_{\Stab}(\sket{e^{-iHt}}) \leq e^{t g^* h^*}.
$$
\end{claim}
\begin{proof}
Using Cauchy-Schwartz 
$$
\sum_{j,k = 1}^R   |h^{(\ell)}_{j,k}| \leq R \cdot \|{\bf h}^{(\ell)}\|_2.
$$
Using H\"older's inequality
$$
\|{\bf h}^{(\ell)}\|_2 \leq  \|{\bf h}\|_2 \cdot\| ({\bf Gh})^{\ell -1}\|_{\mathrm{op}} \leq \|{\bf h}\|_2 \cdot ( \|{\bf G}\|_{\rm op}  \|{\bf h}\|_{\rm op})^{\ell -1}\leq \sqrt{R} h^* (g^* h^*)^{\ell - 1}, \quad \forall \ell \geq 1,
$$
where we have used $\norm{h}_2 \leq \sqrt{R} \norm{h}_{\op} = \sqrt{R} h^*$ and the definitions of $h^*,g^*$ in the last inequality. We thus have
$$
\sum_{j,k=1}^R |h_{j,k}^{(\ell)}| \leq R^{3/2} h^* (g^* h^*)^{\ell - 1}.
$$
Substituting the above equation into Eq.~\eqref{eq:interim1_stab_extent_Ut} then gives us
\begin{align}
\xi_{\Stab}(\sket{e^{-iHt}}) \leq 1 + \frac{R^{3/2}}{\sqrt{2^n}} \sum_{\ell = 1}^\infty \frac{|t|^\ell}{\ell !} h^* (g^* h^*)^{\ell - 1}  
&= 1 + \frac{R^{3/2}}{\sqrt{2^n} g^*} \sum_{\ell = 1}^\infty \frac{(|t| g^* h^*)^{\ell}}{\ell !} \\
&\leq 1 + \frac{R^{3/2}}{\sqrt{2^n} g^*} (e^{t g^* h^*} - 1) \\
&\leq e^{t g^* h^*},
\end{align}
where we used the fact that $R \leq 2^{n/3}$ in the last line along with the fact that $g^* \geq 1$ since $g^* = \norm{\bf G}_{\op} \geq \Tr(\mathbf{G})/R = 1$.
\end{proof}
Let $\varepsilon_1 \in (0,1)$ be an error parameter to be fixed later. Moreover, let us choose $t$ such that $t \leq 1/(g^* h^*)$ to ensure that $\xi_{\Stab}(\sket{e^{-iHt})} \leq e$ (Claim~\ref{claim:stab-extent-bound-Ut}). We now prepare copies of the Choi state $\sket{e^{- i Ht}}$ and invoke \Cref{corr:bounded-stab-extent-learning} (with error instantiated as $\varepsilon_1/\sqrt{2}$ and stabilizer extent as $e$ there) deduce: There exists an algorithm with $\operatorname{poly} (n (1/\varepsilon_1)^{\log(1/\varepsilon_1)}\log (1/\delta))$ sample and time complexity that with probability $\geq 1-\delta$ learns a $O(1/\varepsilon_1^2)$ stabilizer-rank state $\ket{\psi'}$ such that 
\begin{equation}\label{eq:promise_psiprime}
\|\ket{\psi'} - e^{i \theta} \sket{e^{-iHt}} \| \leq \varepsilon_1,    
\end{equation}
for some $\theta \in [0,2\pi)$. We now output an estimate $\widehat{H}$ of $H$ as follows. Let us define $z := \la \EPR_n | \psi'\ra$~\footnote{This can be computed efficiently in $O(n^3/\varepsilon_1^2)$ classical time using the stabilizer inner product algorithm of \cite{garcia2017geometry} as $\ket{\EPR_n}$ is a stabilizer state and $\ket{\psi'}$ is a $O(1/\varepsilon_1^2)$ stabilizer rank state.} and set $\phi := \arg z$. We then define $\ket{\widehat{\psi}} := e^{-i \phi} \ket{\psi'}$ and set 
$$
\sket{\widetilde{H}} := \frac{\ket{\widehat{\psi}} - \ket{\EPR_n}}{-it}.
$$
Let $\mathcal{T}$ be the partial transpose mapping $\ket{A} \ket {B} \mapsto \ket{A}\bra{B^*}$. Then equivalently, we have
\begin{equation}\label{def:sket_tildeH}
\widetilde{H} = \sqrt{2^n} it \cdot \mathcal{T} (\ket{\widehat{\psi}} - \ket{\EPR_n}).    
\end{equation}
and then set $\widehat{H} = (\widetilde{H} + \widetilde{H}^\dagger)/2$ to ensure the output Hamiltonian is Hermitian. 

We now argue that $\widehat{H}$ is indeed close to $H$ for an appropriate choice of $t$. We first observe
\begin{equation}\label{eq:taylor_expansion_sket_Ut2}
\sket{e^{- i Ht}} = \ket{\EPR_n} - i t \sket{H} + \underbrace{\sum_{k = 2}^{\infty}\frac{(-it)^{k}}{k!} \sket{H^k}}_{:=R_t},    
\end{equation}
where $R_t$ is defined as indicated. Directly evaluating
\begin{align}
\norm{\widehat{H} - H}_{\overline 2} = \norm{\frac{\widetilde{H} + \widetilde{H}^\dagger}{2} - \frac{H + H^\dagger}{2}}_{\overline 2} 
& \leq \norm{\widetilde{H} - H}_{\overline 2} \\
&= \norm{\sket{\widetilde{H}} - \sket{H}} \\
&= \norm{ \frac{\ket{\EPR_n} - \ket{\widehat{\psi}}}{it} - \frac{\ket{\EPR_n} - \sket{e^{-itH}} + R_t}{it}} \\
&\leq \frac{\| \ket{\widehat{\psi}} - \sket{e^{-itH}} \| + \|R_t\|}{t}, \label{eq:promise_hatH2}
\end{align}
where we used $H = H^\dagger$ in the first equality, $\norm{\widetilde{H}^\dagger - H^\dagger} = \norm{\widetilde{H} - H}$ along with the triangle inequality in the second inequality, the definition of $\sket{\widetilde{H}}$ (Eq.~\eqref{def:sket_tildeH}) in the third line along with the Taylor expansion of $\sket{e^{-itH}}$ (Eq.~\eqref{eq:taylor_expansion_sket_Ut2}), and finally the triangle inequality in the last line. We now bound $R_t$ as
\begin{equation}\label{eq:norm_bound_Rt2}
\norm{R_t} = \left\|\sum_{k = 2}^{\infty}\frac{(-it)^{k}}{k!} \sket{H^k}\right\| \leq \sum_{k = 2}^{\infty}\frac{t^{k}}{k!} \left\|\sket{H^k}\right\|.    
\end{equation}
We note that
$$
\left\|\sket{H^\ell}\right\| \leq \frac{1}{\sqrt{2^n}} \sum_{j,k} |{\bf h}^{(\ell)}_{j,k}| \leq \frac{R}{\sqrt{2^n}} \cdot \|{\bf h}\|_2 \cdot (\norm{{\bf G}}_{\rm op} \norm{{\bf h}}_{\rm op})^{\ell -1} \leq \frac{R^{3/2}}{\sqrt{2^n}} h^* (g^* h^*)^{\ell -1} \leq h^* (g^* h^*)^{\ell -1}, \quad \forall \ell \geq 1,
$$
where we have used $\norm{h}_2 \leq \sqrt{R} \norm{h}_{\op} = \sqrt{R} h^*$ along with the definitions of $h^*,g^*$ in the third inequality and $R \leq 2^{n/3}$ in the last inequality. Substituting the above equation into Eq.~\eqref{eq:norm_bound_Rt2} gives us
\begin{align}\label{eq:ub_Rt2}
\norm{R_t} \leq \frac{1}{g^*} \sum_{k = 2}^{\infty}\frac{(t g^* h^*)^{k}}{k!} = e^{t g^* h^*} - 1 - t g^* h^* \leq \frac{(t g^* h^*)^2}{2} e^{t g^* h^*} \leq \frac{(t g^* h^*)^2}{2} e,
\end{align}
where we used $g^* = \norm{\mathbf{G}}_{\op} \geq \Tr(\mathbf{G})/R = 1$ in the second inequality, $e^x - 1 - x \leq (x^2/2)e^x, \forall x \geq 0$ in the third inequality and used the fact that $t\leq 1/(g^* h^*)$ in the last inequality. 

We now upper bound $\|\ket{\widehat{\psi}} - \sket{e^{-itH}}\|$ similarly to what we had done in the proof of Theorem~\ref{thm:HL_low_clifford_extent}. We first note that 
\begin{align}
\|\ket{\widehat{\psi}} - \sket{e^{-itH}}\| = \| e^{-i \phi} \ket{\psi'} - \sket{e^{-itH}}\| &=   \| e^{-i \phi} (\ket{\phi'} - e^{i \theta} \sket{e^{-itH}}) + (e^{i(\theta - \phi)} - 1)\sket{e^{-itH}} \| \nonumber \\
&\leq   \| \ket{\psi'} - e^{i \theta} \sket{e^{-itH}} \| + |e^{i(\theta - \phi)} - 1| \cdot \|\sket{e^{-itH}} \| \nonumber \\
&\leq  \varepsilon_1 + |e^{i(\theta - \phi)} - 1|, \label{eq:bound_hatpsi}
\end{align}
where we used Eq.~\eqref{eq:promise_psiprime} in the third line along with the fact that $\sket{e^{-itH}}=1$. To bound the second term in Eq.~\eqref{eq:bound_hatpsi}, let us first define $z:= \la \EPR_n | \psi' \ra$ and then $\phi:= \arg z$. We then note that
\begin{align}
|z - e^{i \theta}| \leq |z-e^{i \theta} \la \EPR_n \sket{e^{-itH}}| + |\la \EPR_n \sket{e^{-itH}} - 1| 
&\leq \|\ket{\psi'} - e^{i\theta} \sket{e^{-itH}}| + |\la \EPR_n \sket{e^{-itH}} - 1| \\
\leq \varepsilon_1 + |\la \EPR_n \sket{e^{-itH}} - 1|, \label{eq:error_z2}
\end{align}
where we used Eq.~\eqref{eq:promise_psiprime} in the second line. We can use Eq.~\eqref{eq:taylor_expansion_sket_Ut2} to bound the second term above by first noting that
$$
\la \EPR_n \sket{e^{-itH}} = \la \EPR_n | \Big(\ket{\EPR_n} - it \sket{H} + R_t \Big) = 1 - it \frac{\Tr(H)}{N} + \la \EPR_n | R_t \ra =  1 +  \la \EPR_n | R_t \ra 
$$
since $\Tr(H)=0$ which implies
\begin{equation}
|\la \EPR_n \sket{e^{-itH}} - 1| \leq \norm{R_t}.
\end{equation}
Substituting the above equation into Eq.~\eqref{eq:error_z2} gives us
\begin{equation}\label{eq:error_z3}
|z - e^{i \theta}| \leq \varepsilon_1 + \norm{R_t}    
\end{equation}
Let $\Delta := \varepsilon_1 + \norm{R_t}$. We now note that
\begin{align}
|e^{i(\theta - \phi)} - 1| = |e^{i\phi} - e^{i\theta}| \leq |e^{i \phi} - z| + |z - e^{i \theta}| 
&= \Big|\frac{z}{|z|} - z\Big| +  \Delta \\
&\leq \Big||z| - 1\Big| + \Delta \\
&\leq |z - e^{i \theta}| + \Delta \\
&\leq 2 \Delta, \label{eq:bound_term3}
\end{align}
where we used the definition of $\phi = \arg z$ in the third inequality in the first line along with Eq.~\eqref{eq:error_z3}, used $| |a| - |b| | \leq |a-b|$ for complex numbers $a,b$ (with $a=z$ and $b=e^{i\theta}$) in the third line, and finally Eq.~\eqref{eq:error_z3} in the last line. Substituting Eq.~\eqref{eq:bound_term3} into Eq.~\eqref{eq:bound_hatpsi} gives us
$$
\|\ket{\widehat{\psi}} - \sket{e^{-itH}}\| \leq 3 \varepsilon_1 + 2 \norm{R_t}.
$$
Substituting the above equation and the bound on $\norm{R_t}$ from Eq.~\eqref{eq:ub_Rt2} back into Eq.~\eqref{eq:promise_hatH2} gives us
$$
\norm{\widehat{H} - H}_{\overline 2} \leq \frac{3 \varepsilon_1}{t} + \frac{3 t (g^* h^*)^2 e}{2}.
$$
Setting $t = \varepsilon/(3 e (g^* h^*)^2)$ and $\varepsilon_1 = \varepsilon t/6 = \varepsilon^2/(18 e (g^* h^*)^2)$ gives us the desired result of 
$$
\norm{\widehat{H} - H}_{\overline 2} \leq \varepsilon.
$$
The query complexity is due to application of Corollary~\ref{corr:bounded-stab-extent-learning} to learn $\ket{\psi'}$ which is $\poly(n (1/\varepsilon_1)^{\log(1/\varepsilon_1)}) = \poly(n (g^* h^*/\varepsilon)^{\log(g^* h^*/\varepsilon)})$. The total time evolution is then $\poly(n (g^* h^*/\varepsilon)^{\log(g^* h^*/\varepsilon)})$. The time complexity associated with using Corollary~\ref{cor:learning_low_clifford_extent} is $\poly(n (g^* h^*/\varepsilon)^{\log(g^* h^*/\varepsilon)})$. The classical time complexity in computing $\arg z$ is $O(n^3/\varepsilon_1^2) = \poly(n g^* h^*/\varepsilon)$ as we remarked earlier (using the classical algorithm to compute stabilizer inner products from \cite{garcia2017geometry}). The overall time complexity of the algorithm is $\poly(n (g^* h^*/\varepsilon)^{\log(g^* h^*/\varepsilon)})$. This completes the proof.
\end{proof}

\subsubsection{Examples of non-Pauli-sparse Hamiltonians that are Clifford sparse}
We end this section with giving examples of families of Hamiltonians that have substantially lower sparsity in the Clifford basis compared with that in the Pauli basis. 

\paragraph{Exchange interactions:} For two disjoint subsets $S_1, S_2$ of $n$ qubits with $|S_1| = |S_2| = k$, define the exchange interaction $E_{S_1, S_2}$ as the operation
$$
E_{S_1, S_2} : \ket{x} \mapsto \ket{\sigma_{S_1, S_2} \circ x},
$$
where $\sigma_{S_1, S_2}$ is the permutation swapping subsets $S_1$ and $S_2$. We can write 
$$
E_{S_1, S_2} = 1/2^k \sum_{P \in \{I, X, Y,Z\}^k} P_{S_1} \otimes P_{S_2},
$$
which is $2^k$ sparse in the Pauli basis. Now consider a Hamiltonian of the form 
$$
H = \sum_{j = 1}^m \alpha_j E_j,
$$
where each $E_j$ is an exchange interaction between subsets of size $k$. In the Pauli basis $H$ can be as dense as $m \cdot 4^k$, however in the Clifford basis it can be written as a sum of $m$ Clifford elements. Nevertheless the Clifford extent (and as a matter of fact the Pauli extent) of this operation is $\leq \sum_j |\alpha_j|$.

\paragraph{Sum of mutually commuting Pauli sets:} Consider family of Hamiltonians of the form
$$
H = \alpha_1 \sum_{P \in \mathcal P_1} P + \alpha_2 \sum_{Q \in \mathcal P_2} Q,
$$
where $\mathcal{P}_1$ and $\mathcal{P}_2$ are each abelian Pauli subgroups of size $2^{m_1}$ and $2^{m_2}$, respectively. We can then write
$$
H = \alpha_1 C_1 \left(\sum_{y \in \{0,1\}^{m_1} 0^{n - m_1}} Z^{y}\right) C_1^\dagger + \alpha_2 C_2 \left(\sum_{y \in \{0,1\}^{m_1} 0^{n - m_2}} Z^{y}\right) C_2^\dagger,
$$
where $C_1$ and $C_2$ are respectively the Clifford unitaries diagonalizing the symplectic matrix representing $\mathcal{P}_1$ and $\mathcal{P}_2$. Then
$$
H = \alpha_1 2^{m_1} C_1 (\ket{0^{m_1}}\bra{0^{m_1}} \otimes I^{n - m_1})C_1^\dagger + \alpha_2 2^{m_2}C_2 (\ket{0^{m_2}}\bra{0^{m_2}} \otimes I^{n -m_2})C_2^\dagger. 
$$
Therefore if we can approximate $\ket{0^m}\bra{0^m}$ as a sum of $r_m (\delta)$ Clifford unitaries within error $\delta$ then $H$ can be approximated within error $\delta$ as a sum of $r_{m_1} (\frac{\delta}{2|\alpha_1| 2^{m_1}}) + r_{m_2} (\frac{\delta}{2|\alpha_2| 2^{m_2}})$. We can naively expand $\ket{0^m}\bra{0^m}$ as a sum of $2^m$ Pauli elements. This bound is tight if we restrict to Pauli expansions and $\delta = 0$. However, we can show that $r_m(0) \leq 2^{m/2}$. For simplicity assume $m$ is even. To see this note that 
$\ket{00}\bra{00} = X^{\otimes 2} \frac{(I - CZ)}2 X^{\otimes 2}$ which is a sum of two Clifford operations. Therefore we can group $\ket{0^m}\bra{0^m} = (\ket{00}\bra{00})^{m/2}$ and expand each group in terms of $2$ Clifford unitaries. However if we are intrested in the Clifford extent we note that the Clifford extent of computational basis outer products such as $\ket{x}\bra{y}$ is equal to $1$. To see this note that $\ket{x}\bra{y}$ is Clifford equivalent to $\ket{0^n}\bra{0^n}$. First note for any decomposition $\ket{0^n}\bra{0^n} = \sum_j c_j C_j$ into Clifford operations
$$
1 = \|\ket{0^n}\bra{0^n}\|_{\rm op} \leq 
\sum_{j} |c_j|\cdot \|C_j\|_{\rm op} \leq \xi_{\Cliff}(\ket{0^n}\bra{0^n})
$$
For the upper bound consider the decomposition $1/2^n\sum_{S \subseteq\{0,1\}^n} Z^S$. Therefore the Clifford extent of $H$ is at most $|\alpha_1| 2^{m_1} + |\alpha_2| 2^{m_2}$.

We note that $\sket{H}$ has stabilizer rank at most $2$. The family described in \Cref{eq:Hamiltonian-family} can furthermore be described as sum of rank $1$ matrices of the form $ C\ket{0}\bra{0} D$ for Clifford $C$ and $D$.

\paragraph{Sparsity in the computational basis:} Consider families of Hamiltonians that are $K$-sparse in the computational basis, i.e., the number of nonzero elements in each row or column of the Hamiltonian in the computational basis is at most $K$. Sometimes this definition of sparsity is used in tasks such as Hamiltonian simulation in the black-box setting. We can also view this family as a special case of the family specified by \Cref{eq:Hamiltonian-family}.
 Consider a Hamiltonian
$$
H = \sum_{x,y \in \{0,1\}^n} h_{x,y} \ket{x}\bra{y} 
$$
from this family. We note that using row/column sparsity we can write 
$$
H = \sum_{x \in \{0,1\}^n} \ket{x}\bra{h(x)}, \qquad \ket{h(x)} := \sum_{y \in \{0,1\}^n} h^*_{xy}\ket{y}
$$
with the assumption that
$$
\|h(x)\|_0 \leq K.
$$

This Hamiltonian in the Pauli basis can be exponentially dense. For instance, to decompose $H = \ket{0^n}\bra{0^n}$ in the Pauli basis we need $2^n$ terms. In general we can consider examples with sparsity $\Omega (4^n)$. We can however obtain fewer size decompositions in the Clifford basis. In particular 
$$
\ket{x}\bra{y} = X^x \ket{0^n}\bra{0^n}X^y,
$$
where $X^x = \prod_j X_j^{x_j}$ for $x\in \{0,1\}^n$. Therefore we can approximate each term using $r_n(\delta)$ Clifford unitaries as defined in the previous example. This implies an upper bound of $2^{3n/2} \cdot K$.

As discussed previously the Clifford rank of $\ket{x}\bra{y}$ is $1$ therefore, the Clifford rank of $H$ is at most $\sum_{x,y} |h_{x,y}|$.

\bibliographystyle{alpha}
\bibliography{references}

@inproceedings{hinsche2025clifford,
author = {Hinsche, Marcel and Bao, Zongbo and van Dordrecht, Philippe and Eisert, Jens and Bri\"{e}t, Jop and Helsen, Jonas},
title = {Clifford Testing: Algorithms and Lower Bounds},
year = {2026},
isbn = {9798400725364},
publisher = {Association for Computing Machinery},
address = {New York, NY, USA},
url = {https://doi.org/10.1145/3798129.3800801},
doi = {10.1145/3798129.3800801},
booktitle = {Proceedings of the 58th Annual ACM Symposium on Theory of Computing},
pages = {869–873},
numpages = {5},
location = {Salt Lake City, UT, USA},
series = {STOC '26}
}

@inproceedings{chen2025stabilizer,
author = {Chen, Sitan and Gong, Weiyuan and Ye, Qi and Zhang, Zhihan},
title = {{Stabilizer Bootstrapping: A Recipe for Efficient Agnostic Tomography and Magic Estimation}},
year = {2025},
isbn = {9798400715105},
publisher = {Association for Computing Machinery},
address = {New York, NY, USA},
url = {https://doi.org/10.1145/3717823.3718191},
doi = {10.1145/3717823.3718191},
booktitle = {Proceedings of the 57th Annual ACM Symposium on Theory of Computing},
pages = {429–438},
numpages = {10},
location = {Prague, Czechia},
series = {STOC '25}
}

@article{wadhwa2025apt,
  title = {Agnostic {P}rocess {T}omography},
  author = {Wadhwa, Chirag and Lewis, Laura and Kashefi, Elham and Doosti, Mina},
  journal = {PRX Quantum},
  volume = {6},
  issue = {4},
  pages = {040371},
  numpages = {54},
  year = {2025},
  month = {Dec},
  publisher = {American Physical Society},
  doi = {10.1103/q2nb-zg9m},
  url = {https://link.aps.org/doi/10.1103/q2nb-zg9m}
}

@inproceedings{buadescu2021qda,
author = {B\u{a}descu, Costin and O'Donnell, Ryan},
title = {Improved Quantum data analysis},
year = {2021},
isbn = {9781450380539},
publisher = {Association for Computing Machinery},
address = {New York, NY, USA},
url = {https://doi.org/10.1145/3406325.3451109},
doi = {10.1145/3406325.3451109},
booktitle = {Proceedings of the 53rd Annual ACM SIGACT Symposium on Theory of Computing},
pages = {1398–1411},
numpages = {14},
location = {Virtual, Italy},
series = {STOC 2021}
}

@article{dong2025linear,
  title={Linear-Size QAC0 Channels: Learning, Testing and Hardness},
  author={Dong, Yangjing and Ou, Fengning and Yao, Penghui},
  journal={arXiv preprint arXiv:2510.00593},
  year={2025}
}

@article{flammia2011direct,
  title         = {Direct Fidelity Estimation from Few {Pauli} Measurements},
  author        = {Flammia, Steven T. and Liu, Yi-Kai},
  journal       = {Physical Review Letters},
  volume        = {106},
  pages         = {230501},
  year          = {2011},
  doi           = {10.1103/PhysRevLett.106.230501},
  eprint        = {1104.4695},
  archivePrefix = {arXiv},
  primaryClass  = {quant-ph}
}

@article{dasilva2011practical,
  title         = {Practical Characterization of Quantum Devices without Tomography},
  author        = {da Silva, Marcus P. and Landon-Cardinal, Olivier and Poulin, David},
  journal       = {Physical Review Letters},
  volume        = {107},
  pages         = {210404},
  year          = {2011},
  doi           = {10.1103/PhysRevLett.107.210404},
  eprint        = {1104.3835},
  archivePrefix = {arXiv},
  primaryClass  = {quant-ph}
}

@article{magesan2011scalable,
  title         = {Scalable and Robust Randomized Benchmarking of Quantum Processes},
  author        = {Magesan, Easwar and Gambetta, J. M. and Emerson, Joseph},
  journal       = {Physical Review Letters},
  volume        = {106},
  pages         = {180504},
  year          = {2011},
  doi           = {10.1103/PhysRevLett.106.180504},
  eprint        = {1009.3639},
  archivePrefix = {arXiv},
  primaryClass  = {quant-ph}
}

@article{temme2017error,
  title         = {Error Mitigation for Short-Depth Quantum Circuits},
  author        = {Temme, Kristan and Bravyi, Sergey and Gambetta, Jay M.},
  journal       = {Physical Review Letters},
  volume        = {119},
  pages         = {180509},
  year          = {2017},
  doi           = {10.1103/PhysRevLett.119.180509},
  eprint        = {1612.02058},
  archivePrefix = {arXiv},
  primaryClass  = {quant-ph}
}

@article{endo2018practical,
  title         = {Practical Quantum Error Mitigation for Near-Future Applications},
  author        = {Endo, Suguru and Benjamin, Simon C. and Li, Ying},
  journal       = {Physical Review X},
  volume        = {8},
  pages         = {031027},
  year          = {2018},
  doi           = {10.1103/PhysRevX.8.031027},
  eprint        = {1712.09271},
  archivePrefix = {arXiv},
  primaryClass  = {quant-ph}
}

@article{czarnik2021error,
  title         = {Error Mitigation with {Clifford} Quantum-Circuit Data},
  author        = {Czarnik, Piotr and Arrasmith, Andrew and Coles, Patrick J. and Cincio, Lukasz},
  journal       = {Quantum},
  volume        = {5},
  pages         = {592},
  year          = {2021},
  doi           = {10.22331/q-2021-11-26-592},
  eprint        = {2005.10189},
  archivePrefix = {arXiv},
  primaryClass  = {quant-ph}
}

@article{lowe2021unified,
  title         = {Unified Approach to Data-Driven Quantum Error Mitigation},
  author        = {Lowe, Angus and Gordon, Max Hunter and Czarnik, Piotr and Arrasmith, Andrew and Coles, Patrick J. and Cincio, Lukasz},
  journal       = {Physical Review Research},
  volume        = {3},
  pages         = {033098},
  year          = {2021},
  doi           = {10.1103/PhysRevResearch.3.033098},
  eprint        = {2011.01157},
  archivePrefix = {arXiv},
  primaryClass  = {quant-ph}
}

@article{bravyi2019simulation,
  title         = {Simulation of Quantum Circuits by Low-Rank Stabilizer Decompositions},
  author        = {Bravyi, Sergey and Browne, Dan and Calpin, Padraic and Campbell, Earl and Gosset, David and Howard, Mark},
  journal       = {Quantum},
  volume        = {3},
  pages         = {181},
  year          = {2019},
  doi           = {10.22331/q-2019-09-02-181},
  eprint        = {1808.00128},
  archivePrefix = {arXiv},
  primaryClass  = {quant-ph}
}

@article{filmus2014real,
  title={Real analysis in computer science: A collection of open problems},
  author={Filmus, Yuval and Hatami, Hamed and Heilman, Steven and Mossel, Elchanan and O’Donnell, Ryan and Sachdeva, Sushant and Wan, Andrew and Wimmer, Karl},
  journal={Preprint available at https://simons. berkeley. edu/sites/default/files/openprobsmerged. pdf},
  year={2014}
}

@article{fattal2004entanglement,
  title={Entanglement in the stabilizer formalism},
  author={Fattal, David and Cubitt, Toby S and Yamamoto, Yoshihisa and Bravyi, Sergey and Chuang, Isaac L},
  journal={arXiv preprint quant-ph/0406168},
  year={2004}
}

@article{grewal2025query,
  title={Query-optimal estimation of unitary channels via {P}auli dimensionality},
  author={Grewal, Sabee and Liang, Daniel},
  journal={arXiv preprint arXiv:2510.00168},
  year={2025}
}

@article{arunachalam2026tomography,
  title={Tomography of quantum states with bounded extent},
  author={Arunachalam, Srinivasan and Dutt, Arkopal},
  journal={arXiv preprint arXiv:2606.07425},
  year={2026}
}

@article{low2009learning,
  title={Learning and testing algorithms for the Clifford group},
  author={Low, Richard A},
  journal={Physical Review A—Atomic, Molecular, and Optical Physics},
  volume={80},
  number={5},
  pages={052314},
  year={2009},
  publisher={APS}
}

@InProceedings{arunachalam2024learning,
  author =	{Arunachalam, Srinivasan and Dutt, Arkopal and Escudero Guti\'{e}rrez, Francisco and Palazuelos, Carlos},
  title =	{{Learning Low-Degree Quantum Objects}},
  booktitle =	{51st International Colloquium on Automata, Languages, and Programming (ICALP 2024)},
  pages =	{13:1--13:19},
  series =	{Leibniz International Proceedings in Informatics (LIPIcs)},
  ISBN =	{978-3-95977-322-5},
  ISSN =	{1868-8969},
  year =	{2024},
  volume =	{297},
  editor =	{Bringmann, Karl and Grohe, Martin and Puppis, Gabriele and Svensson, Ola},
  publisher =	{Schloss Dagstuhl -- Leibniz-Zentrum f{\"u}r Informatik},
  address =	{Dagstuhl, Germany},
  URL =		{https://drops.dagstuhl.de/entities/document/10.4230/LIPIcs.ICALP.2024.13},
  URN =		{urn:nbn:de:0030-drops-201563},
  doi =		{10.4230/LIPIcs.ICALP.2024.13}
}

@article{lai2022learning,
  title={Learning quantum circuits of some T gates},
  author={Lai, Ching-Yi and Cheng, Hao-Chung},
  journal={IEEE Transactions on Information Theory},
  volume={68},
  number={6},
  pages={3951--3964},
  year={2022},
  publisher={IEEE}
}

@inproceedings{bao2023testing,
  title={On testing and learning quantum junta channels},
  author={Bao, Zongbo and Yao, Penghui},
  booktitle={The Thirty Sixth Annual Conference on Learning Theory},
  pages={1064--1094},
  year={2023},
  organization={PMLR}
}

@article{honjani2026query,
  title={Query Learning Nearly Pauli Sparse Unitaries in Diamond Distance},
  author={Honjani, Zahra and Heidari, Mohsen},
  journal={arXiv preprint arXiv:2604.00203},
  year={2026}
}

@inproceedings{arunachalam2025testing,
author = {Arunachalam, Srinivasan and Dutt, Arkopal and Escudero Guti\'{e}rrez, Francisco},
title = {Testing and Learning Structured Quantum Hamiltonians},
year = {2025},
isbn = {9798400715105},
publisher = {Association for Computing Machinery},
address = {New York, NY, USA},
url = {https://doi.org/10.1145/3717823.3718289},
doi = {10.1145/3717823.3718289},
booktitle = {Proceedings of the 57th Annual ACM Symposium on Theory of Computing},
pages = {1263–1270},
numpages = {8},
location = {Prague, Czechia},
series = {STOC '25}
}

@article{gottesman1998heisenberg,
  title={The Heisenberg representation of quantum computers},
  author={Gottesman, Daniel},
  journal={arXiv preprint quant-ph/9807006},
  year={1998}
}

@article{bravyi2016improved,
  title={Improved classical simulation of quantum circuits dominated by Clifford gates},
  author={Bravyi, Sergey and Gosset, David},
  journal={Physical review letters},
  volume={116},
  number={25},
  pages={250501},
  year={2016},
  publisher={APS}
}

@article{bu2025quantum,
  title={Quantum higher-order Fourier analysis and the Clifford hierarchy},
  author={Bu, Kaifeng and Gu, Weichen and Jaffe, Arthur},
  journal={Proceedings of the National Academy of Sciences},
  volume={122},
  number={45},
  pages={e2515667122},
  year={2025},
  publisher={National Academy of Sciences}
}

@article{Gross2017SchurWeylDF,
  title={Schur–Weyl Duality for the Clifford Group with Applications: Property Testing, a Robust Hudson Theorem, and de Finetti Representations},
  author={David Gross and Sepehr Nezami and Michael Walter},
  journal={Communications in Mathematical Physics},
  year={2017},
  volume={385},
  pages={1325 - 1393},
  url={https://api.semanticscholar.org/CorpusID:73551109}
}

@article{huang2020predicting,
  title={Predicting many properties of a quantum system from very few measurements},
  author={Huang, Hsin-Yuan and Kueng, Richard and Preskill, John},
  journal={Nature Physics},
  volume={16},
  number={10},
  pages={1050--1057},
  year={2020},
  publisher={Nature Publishing Group UK London}
}

@article{ma2024learning,
  title={Learning $ k $-body Hamiltonians via compressed sensing},
  author={Ma, Muzhou and Flammia, Steven T and Preskill, John and Tong, Yu},
  journal={arXiv preprint arXiv:2410.18928},
  year={2024}
}

@article{yu2023robust,
  title={Robust and efficient Hamiltonian learning},
  author={Yu, Wenjun and Sun, Jinzhao and Han, Zeyao and Yuan, Xiao},
  journal={Quantum},
  volume={7},
  pages={1045},
  year={2023},
  publisher={Verein zur F{\"o}rderung des Open Access Publizierens in den Quantenwissenschaften}
}

@article{dehaene2003clifford,
  title = {Clifford group, stabilizer states, and linear and quadratic operations over {GF(2)}},
  author = {Dehaene, Jeroen and De Moor, Bart},
  journal = {Phys. Rev. A},
  volume = {68},
  issue = {4},
  pages = {042318},
  numpages = {10},
  year = {2003},
  publisher = {American Physical Society},
  doi = {10.1103/PhysRevA.68.042318},
  url = {https://link.aps.org/doi/10.1103/PhysRevA.68.042318}
}

@article{patel2003efficient,
  title={Efficient synthesis of linear reversible circuits},
  author={Patel, Ketan N and Markov, Igor L and Hayes, John P},
  journal={arXiv  quant-ph/0302002},
  year={2003}
}

@inbook{chen2023juntas,
author = {Thomas Chen and Shivam Nadimpalli and Henry Yuen},
title = {Testing and Learning Quantum Juntas Nearly Optimally},
booktitle = {Proceedings of the 2023 Annual ACM-SIAM Symposium on Discrete Algorithms (SODA)},
chapter = {},
pages = {1163-1185},
doi = {10.1137/1.9781611977554.ch43},
URL = {https://epubs.siam.org/doi/abs/10.1137/1.9781611977554.ch43},
eprint = {https://epubs.siam.org/doi/pdf/10.1137/1.9781611977554.ch43}
}

@inproceedings{briet2026near,
title={A near-optimal quadratic Goldreich-Levin algorithm},
author={Bri{\"e}t, Jop and Castro-Silva, Davi},
booktitle={Proceedings of the 2026 Annual ACM-SIAM Symposium on Discrete Algorithms (SODA)},
pages={6233--6239},
year={2026},
organization={SIAM},
doi = {10.1137/1.9781611978971.224},
URL = {https://epubs.siam.org/doi/abs/10.1137/1.9781611978971.224},
eprint = {https://epubs.siam.org/doi/pdf/10.1137/1.9781611978971.224}
}

@article{fanizza2025efficient,
  title={Efficient learning of bosonic {G}aussian unitaries},
  author={Fanizza, Marco and Iyer, Vishnu and Lee, Junseo and Mele, Antonio A and Mele, Francesco A},
  journal={arXiv preprint arXiv:2510.05531},
  year={2025}
}

@article{iyer2025mildly,
  title={Mildly-interacting fermionic unitaries are efficiently learnable},
  author={Iyer, Vishnu},
  journal={arXiv preprint arXiv:2504.11318},
  year={2025}
}

@inproceedings{huang2024shallow,
author = {Huang, Hsin-Yuan and Liu, Yunchao and Broughton, Michael and Kim, Isaac and Anshu, Anurag and Landau, Zeph and McClean, Jarrod R.},
title = {Learning Shallow Quantum Circuits},
year = {2024},
isbn = {9798400703836},
publisher = {Association for Computing Machinery},
address = {New York, NY, USA},
url = {https://doi.org/10.1145/3618260.3649722},
doi = {10.1145/3618260.3649722},
booktitle = {Proceedings of the 56th Annual ACM Symposium on Theory of Computing},
pages = {1343–1351},
numpages = {9},
location = {Vancouver, BC, Canada},
series = {STOC 2024}
}

@article{zhao2024bounded,
  title = {Learning Quantum States and Unitaries of Bounded Gate Complexity},
  author = {Zhao, Haimeng and Lewis, Laura and Kannan, Ishaan and Quek, Yihui and Huang, Hsin-Yuan and Caro, Matthias C.},
  journal = {PRX Quantum},
  volume = {5},
  issue = {4},
  pages = {040306},
  numpages = {63},
  year = {2024},
  month = {Oct},
  publisher = {American Physical Society},
  doi = {10.1103/PRXQuantum.5.040306},
  url = {https://link.aps.org/doi/10.1103/PRXQuantum.5.040306}
}

@inproceedings{zhao2025learning,
author = {Zhao, Andrew},
title = {Learning the Structure of Any Hamiltonian from Minimal Assumptions},
year = {2025},
isbn = {9798400715105},
publisher = {Association for Computing Machinery},
address = {New York, NY, USA},
url = {https://doi.org/10.1145/3717823.3718115},
doi = {10.1145/3717823.3718115},
booktitle = {Proceedings of the 57th Annual ACM Symposium on Theory of Computing},
pages = {1201–1211},
numpages = {11},
location = {Prague, Czechia},
series = {STOC '25}
}

@article{sinha2025improved,
  title={Improved Hamiltonian learning and sparsity testing through Bell sampling},
  author={Sinha, Savar D and Tong, Yu},
  journal={arXiv preprint arXiv:2509.07937},
  year={2025}
}

@article{shin2026heisenberg,
  title={Heisenberg-limited Hamiltonian learning without short-time control},
  author={Shin, Myeongjin and Lee, Junseo and Oh, Changhun},
  journal={arXiv preprint arXiv:2604.27838},
  year={2026}
}

@article{hu2025ansatz,
  title = {Ansatz-Free Hamiltonian Learning with Heisenberg-Limited Scaling},
  author = {Hu, Hong-Ye and Ma, Muzhou and Gong, Weiyuan and Ye, Qi and Tong, Yu and Flammia, Steven T. and Yelin, Susanne F.},
  journal = {PRX Quantum},
  volume = {6},
  issue = {4},
  pages = {040315},
  numpages = {30},
  year = {2025},
  month = {Oct},
  publisher = {American Physical Society},
  doi = {10.1103/j7b8-pb77},
  url = {https://link.aps.org/doi/10.1103/j7b8-pb77}
}

@article{huang2023HL,
  title = {Learning Many-Body Hamiltonians with Heisenberg-Limited Scaling},
  author = {Huang, Hsin-Yuan and Tong, Yu and Fang, Di and Su, Yuan},
  journal = {Phys. Rev. Lett.},
  volume = {130},
  issue = {20},
  pages = {200403},
  numpages = {7},
  year = {2023},
  month = {May},
  publisher = {American Physical Society},
  doi = {10.1103/PhysRevLett.130.200403},
  url = {https://link.aps.org/doi/10.1103/PhysRevLett.130.200403}
}

@article{dutkiewicz2024advantage,
  title={The advantage of quantum control in many-body Hamiltonian learning},
  author={Dutkiewicz, Alicja and O'Brien, Thomas E and Schuster, Thomas},
  journal={Quantum},
  volume={8},
  pages={1537},
  year={2024},
  publisher={Verein zur F{\"o}rderung des Open Access Publizierens in den Quantenwissenschaften}
}

@article{de2026learning,
  title={Learning Hamiltonians at Long Times},
  author={de Pradenne, Constantin Cedillo Vayson and Cotler, Jordan and Huang, Hsin-Yuan},
  journal={arXiv preprint arXiv:2606.05690},
  year={2026}
}

@article{caro2024learn,
author = {Caro, Matthias C.},
title = {Learning Quantum Processes and Hamiltonians via the Pauli Transfer Matrix},
year = {2024},
issue_date = {June 2024},
publisher = {Association for Computing Machinery},
address = {New York, NY, USA},
volume = {5},
number = {2},
url = {https://doi.org/10.1145/3670418},
doi = {10.1145/3670418},
journal = {ACM Transactions on Quantum Computing},
month = jun,
articleno = {14},
numpages = {53}
}

@article{abbas2025nearly,
  title={Nearly optimal algorithms to learn sparse quantum Hamiltonians in physically motivated distances},
  author={Abbas, Amira and Cerrato, Nunzia and Guti{\'e}rrez, Francisco Escudero and Grinko, Dmitry and Mele, Francesco Anna and Sinha, Pulkit},
  journal={arXiv preprint arXiv:2509.09813},
  year={2025}
}

@INPROCEEDINGS{bakshi2024structure,
  author={Bakshi, Ainesh and Liu, Allen and Moitra, Ankur and Tang, Ewin},
  booktitle={2024 IEEE 65th Annual Symposium on Foundations of Computer Science (FOCS)}, 
  title={Structure Learning of Hamiltonians from Real-Time Evolution}, 
  year={2024},
  volume={},
  number={},
  pages={1037-1050},
  doi={10.1109/FOCS61266.2024.00069}}

@article{castaneda2025hamiltonian,
  title={Hamiltonian learning via shadow tomography of pseudo-choi states},
  author={Castaneda, Juan and Wiebe, Nathan},
  journal={Quantum},
  volume={9},
  pages={1700},
  year={2025},
  publisher={Verein zur F{\"o}rderung des Open Access Publizierens in den Quantenwissenschaften}
}

@InProceedings{abdy2023phase,
  author =	{Arunachalam, Srinivasan and Bravyi, Sergey and Dutt, Arkopal and Yoder, Theodore J.},
  title =	{{Optimal Algorithms for Learning Quantum Phase States}},
  booktitle =	{18th Conference on the Theory of Quantum Computation, Communication and Cryptography (TQC 2023)},
  pages =	{3:1--3:24},
  series =	{Leibniz International Proceedings in Informatics (LIPIcs)},
  ISBN =	{978-3-95977-283-9},
  ISSN =	{1868-8969},
  year =	{2023},
  volume =	{266},
  editor =	{Fawzi, Omar and Walter, Michael},
  publisher =	{Schloss Dagstuhl -- Leibniz-Zentrum f{\"u}r Informatik},
  address =	{Dagstuhl, Germany},
  URL =		{https://drops.dagstuhl.de/entities/document/10.4230/LIPIcs.TQC.2023.3},
  URN =		{urn:nbn:de:0030-drops-183139},
  doi =		{10.4230/LIPIcs.TQC.2023.3}
}

@article{bardenet2015concentration,
  title={Concentration inequalities for sampling without replacement},
  author={Bardenet, R{\'e}mi and Maillard, Odalric-Ambrym},
  year={2015}
}

@article{kalra2026stabilizer,
  title={{Stabilizer Ranks, Barnes Wall Lattices and Magic Monotones}},
  author={Kalra, Amolak Ratan and Sinha, Pulkit},
  journal={Quantum},
  volume={10},
  pages={2179},
  year={2026},
  publisher={Verein zur F{\"o}rderung des Open Access Publizierens in den Quantenwissenschaften}
}

@inproceedings{mehraban2025improved,
  title={Improved bounds for testing low stabilizer complexity states},
  author={Mehraban, Saeed and Tahmasbi, Mehrdad},
  booktitle={Proceedings of the 57th Annual ACM Symposium on Theory of Computing},
  pages={1222--1233},
  year={2025}
}

@inproceedings{mehraban2024quadratic,
  title={Quadratic lower bounds on the approximate stabilizer rank: A probabilistic approach},
  author={Mehraban, Saeed and Tahmasbi, Mehrdad},
  booktitle={Proceedings of the 56th Annual ACM Symposium on Theory of Computing},
  pages={608--619},
  year={2024}
}

@article{koenig2014efficiently,
  title={How to efficiently select an arbitrary Clifford group element},
  author={Koenig, Robert and Smolin, John A},
  journal={Journal of Mathematical Physics},
  volume={55},
  number={12},
  year={2014},
  publisher={AIP Publishing}
}

@article{garcia2017geometry,
  title={On the geometry of stabilizer states},
  author={Garc{\'\i}a, H{\'e}ctor J and Markov, Igor L and Cross, Andrew W},
  journal={arXiv preprint arXiv:1711.07848},
  year={2017}
}

\appendix

\section{Weak agnostic tomography protocol for Clifford unitaries}
\label{sec:wal_appendix}
In this section of the appendix, we will describe the weak agnostic protocol for Clifford unitaries that was referred to in Section~\ref{sec:tech_overview}. In particular, we have the following result.

\begin{theorem}
Let $\tau,\delta \in (0,1)$. Given an unknown $n$-qubit unitary $U$ that has Clifford fidelity $\geq \tau$, there is a quantum algorithm that with success probability $\geq 1-\delta$ outputs a Clifford unitary $V$ such that $|\la\!\la U | V \ra\!\ra|^2 \geq \tau^2/8$. The algorithm uses $\poly(n, (1/\tau)^{\log(1/\tau)}\log(1/\delta))$ many queries to $U$ and time complexity.
\end{theorem}
\begin{proof}
From \cite{hinsche2025clifford}, we know that
$$
\calF_{\calS}(\sket{U}) \geq \calF_{\Cliff(n)}(U) \geq \calF_{\calS}(\sket{U})^6.
$$
From the given promise, we have that $\calF_{\calS}(\sket{U}) \geq \tau$. Using \cite{chen2025stabilizer}, we can learn a $2n$-qubit stabilizer state $\ket{\phi}$ such that 
$$
|\la \phi | U \ra\!\ra|^2 \geq \tau/2.
$$
Since $\ket{\phi}$ is a stabilizer state on the
bipartition $A|B$, with $|A|=|B|=n$, we can compute local Clifford unitaries
$C_A,C_B$ and an integer $k\in\{0,\ldots,n\}$ (using \cite{fattal2004entanglement}) such that
$$
(C_A \otimes C_B) \ket{\phi} = \ket{\EPR_k} \otimes \ket{0}^{n-k} \otimes \ket{0}^{n-k},
$$
where $\ket{\EPR_k}$ is an $k$-pair $\EPR$ state across the $AB$ cut. We now argue that $k \geq n - \log(2/\tau)$.\footnote{Also, this is tight in some sense, since we can find stabilizer states which have fidelity $\geq \tau/2$ and have entanglement entropy $\leq n - \log(2/\tau)$ (maybe up to constants,etc.)} Denote $U' = C_A U C_B^T$ and let $t:=n-k$. We note that
\begin{equation}\label{eq:promise_k}
\frac{\tau}{2} \leq |\la \phi | U \ra\!\ra|^2 = \left|\Big(\bra{\EPR_k} \otimes \bra{0}^{2n-2k} \Big)\sket{U'} \right|^2 \leq 2^{k - n}  \implies \log(\tau/2) \leq k - n \implies k \geq n - \log(2/\tau),
\end{equation}
where in the implication we noted that 
$$
\ket{\EPR_k} = \sqrt{2^{-k}} \sum_{x \in \{0,1\}^k} \ket{x}\ket{x}
$$
and then the last term before the implication is given by
$$
2^{-(n+k)} \Big| \sum_{x\in\{0,1\}^k} \bra{x,0^t} U' \ket{x,0^t} \Big|^2 \leq 2^{-(n+k)} 2^{2k} = 2^{k-n}.
$$

To obtain the desired Clifford unitary, we now consider the following procedure. For each $s\in\{0,1\}^t$, let $W_s := I_{2^k}\otimes Z^s$,  where \(Z^s = Z^{s_1}\otimes\cdots\otimes Z^{s_t}\). Each \(W_s\) is a simple $n$-qubit Clifford unitary. Particularly, $\la\!\la W_s | U' \ra\!\ra = 2^{-n}\Tr(W_s U')$. Averaging over $s\in\{0,1\}^t$, we obtain
$$
\mathbb{E}_{s\in\{0,1\}^t} \la\!\la W_s | U' \ra\!\ra = 2^{-n} \sum_{x\in\{0,1\}^k} \bra{x,0^t}U'\ket{x,0^t}.
$$
Here we used 
$$
\frac{1}{2^t} \sum_{s \in \{0,1\}^t} Z^s = \ket{0^t}\bra{0^t}.
$$
Therefore there exists some $s^\star \in \{0,1\}^t$ such that
$$
|\la\!\la W_{s^\star} | U' \ra\!\ra| \geq 2^{-n} \left| \sum_{s\in\{0,1\}^k} \bra{s,0^t}U'\ket{s,0^t} \right|  =
2^{(k-n)/2} \left| \left( \bra{\EPR_k}_{A_1B_1} \otimes \bra{0}^{\otimes t}_{A_2} \otimes \bra{0}^{\otimes t}_{B_2} \right) \sket{U'} \right|.
$$
Using Eq.~\eqref{eq:promise_k}, we then have
$$
|\la\!\la W_{s^\star} | U' \ra\!\ra|^2 \geq 2^{k-n}\cdot \tau/2 \geq \tau^2/4.
$$
We can then define the Clifford unitary $V := C_A^\dagger W_{s^\star} C_B^\star$ which satisfies
$$
|\la\!\la U | V\ra\!\ra|^2 = |\la\!\la U' | W_{s^\star}\ra\!\ra|^2 \geq \tau^2/4.
$$
So after obtaining \(\ket{\phi}\), we use \cite{fattal2004entanglement} to obtain $k,C_A,C_B$. Since $t \leq \log(2/\tau)$, there are only $2^t \leq 2/\tau$ choices of $s^\star$. We determine $s^\star$ by enumerating through all $s \in \{0,1\}^t$ and estimating the overlap $|\la\!\la U' \mid W_s\ra\!\ra|^2$ using a SWAP test within error $\tau^2/8$, and then setting $s^\star$ as that string which maximizes the fidelity. The final promise would then be
$$
|\la\!\la U | V\ra\!\ra|^2 \geq \tau^2/8.
$$
The complexity of the algorithm is due to using \cite{chen2025stabilizer} which requires $\poly(n,(1/\tau)^{\log(1/\tau)}\log(1/\delta))$ time complexity and determining $s^\star$ which requires $O(1/\tau^5)$ samples and $O(n^2/\tau^5)$ time complexity.
\end{proof}

\section{Exponential separation between $\xi_{\Cliff} (U)$ and $\xi_{\Stab} (\sket{U})$}
\label{app:xi-separations}

A major strategy in this paper to study the decomposition of a unitary $U$ in is to study stabilizer decompositions of the corresponding Choi state $\sket{U}$. A natural question is whether basic parameters of these two objects such as $\xi_{\Cliff} (U)$ and $\xi_{\Stab} (\sket{U})$ are polynomially related to each other. This is particularly important in \Cref{sec:Hamiltonian-learning} where we outlined two different algorithms one relying on the former quantity and another based on the latter quantity for $U = e^{- i Ht}$. Polynomial relationship between the two quantities would imply equivalence between these approaches. We however show that there exist instances of unitary operations for which these two quantities are exponentially separated.

\begin{theorem}
    \label{thm:extent-separation}
    There exists a diagonal $n$ qubit unitary $U$ for which 
    $$
    \xi_{\Stab} (\sket{U}) \leq 3,
    $$
    but
    $$
    \xi_{\Cliff} (U) \geq \Omega (\frac{2^{n/4}}{n}).
    $$  
\end{theorem}

\begin{proof}
    We show that there exists a subset $S$ of size $m = \floor{\sqrt{2^n}}$ of $\{0,1\}^n$ for which the unitary
    $$
    U_S = I - 2 \Pi_S, \quad \Pi_S = \sum_{x \in S} \ket{x}\bra{x} 
    $$
    satisfies the criteria expressed in this theorem.

    We first upper bound $\xi_{\Stab} (\sket{U_S})$. We note that 
    $$
    \sket{U_S} = \frac{1}{\sqrt{2^n}} \sum_{x\in \{0,1\}^n} (-1)^{1_S(x)} \ket{x}\ket{x}
    $$
    which is Clifford equivalent to 
    $$
    \left (\frac{1}{\sqrt{2^n}} \sum_{x\in \{0,1\}^n} (-1)^{1_S(x)} \ket{x}\right ) \otimes \ket{0^n}
    $$
    using $\mathrm{CNOT}$ gates between the two halves. The first factor equals
    $$
    \ket{+^n} - \frac{2}{\sqrt{2^n}}\sum_{x \in S}\ket{x}
    $$
    The stabilizer extent of this state is upper bounded by
    $$
    1 + 2 |S|/\sqrt{2^n} \leq 3
    $$
    for any $S$ of size $m$.

    We now show that there exists a subset $S$ of size $m$ for which the Clifford extent of $U_S$ is at least $\Omega(2^{n/4}/n)$.
    Let $W_S = \Pi_S - \frac{m}{2^n} I$. For any Clifford decomposition
    $$
    U_S = \sum_j c_j C_j
    $$
    we have
    $$
    |\Tr(W_S U_S)| \leq \sum_j |c_j| \max_{C \in \Cliff(n)} |\Tr (C W_S)|
    $$
    As a result
    $$
    \xi_{\Cliff} (U) \geq \frac{|\Tr(W_S U_S)|}{\max_{C \in \Cliff(n)} |\Tr (C W_S)|}
    $$
    We first compute the numerator
    $$
    |\Tr(W_S U_S)| = 2m (1 - \frac{m}{2^n}) \geq m
    $$
    for which we used $m/2^n \leq 1/2$. 

    To find $S$ which puts a suitable upper bound on the denominator we use a probabilistic argument. 
    Let $S$ be a random subset of size $m$. Let $C_x = \bra{x}C\ket{x}$ and let $\bar C = \frac{1}{2^n} \sum_x C_x$. Then 
    $$
    \Tr (C W_S) = \sum_{x \in S} C_x - m \bar C 
    $$
    We now use Hoeffding's inequality (assuming sampling without replacement as in \cite[Proposition 1.2]{bardenet2015concentration}) applied to both imaginary and real parts to obtain
    $$
    \Pr_S (|\sum_{x \in S} C_x - m\bar C| \geq t) \leq 4 e^{- \frac{t^2}{4m}}
    $$
    We note that to obtain the bound above we have applied Hoeffding once to the real part to bound the probability $|\Re (Z_S)| \geq t/\sqrt{2}$, where $Z_S := \Tr (CW_S)$, and another time to bound $|\Im (Z_S)| \geq t/\sqrt{2}$ and used $|Z_S|^2 = |\Im(Z_S)|^2 + |\Re(Z_S)|^2$. 

    We now use the union bound 
    $$
    \Pr_S (\max_C |Z_S| \geq t) \leq 4|\Cliff(n)|e^{- \frac{t^2}{4m}}
    $$
    We know that $N_n := |\Cliff(n)| \leq 2^{2n^2 + 3 n}$ (see for instance \cite[Equation 1]{koenig2014efficiently}). Therefore
    $$
    \Pr_S (\max_C |Z_S| < \sqrt{4 m \ln (4 N_n)} ) > 0
    $$
    which implies that there exists $S$ of size $m$ such that 
    $$
    \max_C |Z_S| < \sqrt{4 m \ln (4 N_n)}
    $$
    hence 
    $$
    \xi_{\Cliff} (U_S) \geq \frac{\sqrt{m}}{\sqrt{4\ln (4 N_n)}} = \Omega (\frac{2^{n/4}}n).
    $$
\end{proof}
\end{document}